\documentstyle[preprint,prd,eqsecnum,aps,epsf]{revtex}

\tightenlines
\begin{document}

    \newcommand{\DSC}{D\hspace{-0.25cm}\slash_{\bot}}
    \newcommand{\DSP}{D\hspace{-0.25cm}\slash_{\|}}
    \newcommand{\DS}{D\hspace{-0.25cm}\slash}
    \newcommand{\DC}{D_{\bot}}
    \newcommand{\DSCX}{D\hspace{-0.20cm}\slash_{\bot}}
    \newcommand{\DSPX}{D\hspace{-0.20cm}\slash_{\|}}
    \newcommand{\DP}{D_{\|}}
    \newcommand{\QV}{Q_v^{+}}
    \newcommand{\QVB}{\bar{Q}_v^{+}}
    \newcommand{\QVP}{Q^{\prime +}_{v^{\prime}} }
    \newcommand{\QVBP}{\bar{Q}^{\prime +}_{v^{\prime}} }
    \newcommand{\QVHZ}{\hat{Q}^{+}_v}
    \newcommand{\QVHZB}{\bar{\hat{Q}}_v{\vspace{-0.3cm}\hspace{-0.2cm}{^{+}} } }
    \newcommand{\QVPHZB}{\bar{\hat{Q}}_{v^{\prime}}{\vspace{-0.3cm}\hspace{-0.2cm}{^{\prime +}}} }

    \newcommand{\QVPHFB}{\bar{\hat{Q}}_{v^{\prime}}{\vspace{-0.3cm}\hspace{-0.2cm}{^{\prime -}} } }
    \newcommand{\QVPHB}{\bar{\hat{Q}}_{v^{\prime}}{\vspace{-0.3cm}\hspace{-0.2cm}{^{\prime}} }   }
    \newcommand{\QVHF}{\hat{Q}^{-}_v}
    \newcommand{\QVHFB}{\bar{\hat{Q}}_v{\vspace{-0.3cm}\hspace{-0.2cm}{^{-}} }}
    \newcommand{\QVH}{\hat{Q}_v}
    \newcommand{\QVHB}{\bar{\hat{Q}}_v}
    \newcommand{\VS}{v\hspace{-0.2cm}\slash}
    \newcommand{\MQ}{m_{Q}}
    \newcommand{\MQP}{m_{Q^{\prime}}}
    \newcommand{\QVHPMB}{\bar{\hat{Q}}_v{\vspace{-0.3cm}\hspace{-0.2cm}{^{\pm}} }}
    \newcommand{\QVHMPB}{\bar{\hat{Q}}_v{\vspace{-0.3cm}\hspace{-0.2cm}{^{\mp}} }  }
    \newcommand{\QVHPM}{\hat{Q}^{\pm}_v}
    \newcommand{\QVHMP}{\hat{Q}^{\mp}_v}

\draft
\title{  Heavy to Light Meson Exclusive Semileptonic Decays \\ in Effective Field
Theory of Heavy Quark}
\author{ W.Y. Wang$\mbox{}^*$, Y.L. Wu$\mbox{}^\dagger$ and M. Zhong
$\mbox{}^\dagger$}
\address{$*$ Department of Physics, Tsinghua University, Beijing 100084, China \\
$\dagger$ Institute of Theoretical Physics, Academia Sinica, 
 Beijing 100080, China }
\maketitle

\begin{abstract}
We present a general study on exclusive semileptonic decays of heavy ($B$, $D$, $B_s$
) to light ($\pi$, $\rho$, $K$, $K^*$) mesons in the framework of effective 
field theory of heavy quark. 
Transition matrix elements of these decays can be systematically characterized 
by a set of wave functions which are independent of the heavy 
quark mass except for the implicit scale dependence. 
Form factors for all these decays are calculated consistently within the effective theory 
framework using the 
light cone sum rule method at the leading order of $1/m_Q$ expansion. 
The branching ratios of these decays are evaluated, and the heavy and light flavor 
symmetry breaking effects are investigated. We also give comparison of 
our results and the predictions from other approaches, among which are the 
relations proposed recently in the framework of large energy effective theory.  

\end{abstract}


\newpage

\section{Introduction}\label{int}
Heavy meson decays, both inclusive and exclusive, have long been an interesting
subject in both experimental and theoretical study. 
These decays play their 
special role in extracting the Cabibbo-Kabayashi-Maskawa (CKM) matrix elements 
and probing new physics beyond 
the standard model (SM). The inclusive decays of heavy mesons are more difficult 
to measure but theoretically cleaner than exclusive decays. 
On the other hand, exclusive decays are cleaner in experimental measurements but 
more difficult in theoretical calculations as they require the knowledge of
form factors, which contains long distance ingredients and has to be estimated
via nonperturbative methods such as sum rules, lattice simulations or phenomenological 
models. 

The heavy to light exclusive decays can be grouped as  
semileptonic decays and rare decays. 
Besides using lattice calculations 
\cite{ape,wuppertal,ukqcd,elc,lat9910021,lat9710057,lat98} and quark models 
\cite{ph0001113,Jaus96,wsb85,ph9807223,isgw2}, 
these decays have been analyzed by using sum rules 
in full QCD \cite{pvesrb,var,ar,arsc,ph9805422,ph9802394,ph0001297,ph9701238,sr9193}. 
Since the heavy meson $B_{(s)}$ or $D_{(s)}$ contains one heavy quark and one light quark, 
it is expected that the heavy quark symmetry (HQS) and the effective field 
theory of heavy quark may help 
to improve our understanding on heavy to light decays. 
Recently, some work has been done in this direction \cite{gzmy,hly,bpi,brho}. 
In particular, in Refs.\cite{bpi,brho} the $B\to \pi (\rho)l\nu$ decays are 
investigated within the framework of 
heavy quark effective field theory (HQEFT) \cite{ylw,wwy,ww}, and the relevant form factors are  
calculated at the leading order of heavy quark expansion, using which $|V_{ub}|$ 
is extracted. 
 
This paper will provide a more general discussion on exclusive heavy to light 
decays within the framework of effective theory. We shall apply the heavy 
quark expansion (HQE) and light cone 
sum rule (LCSR) techniques to more heavy ($B$, $D$, $B_s$) to light ($\pi$, $\rho$, $K$, 
$K^*$) exclusive semileptonic decays. 
Namely, we extend the study to D decays, and also extend it to decays 
into kaon mesons. We know that the reliability of HQE is 
different for B and D mesons, and that the SU(3) symmetry breaking effects 
arise in the kaon systems.  
So, through a consistent study on decays of both bottom and charm mesons 
with the final mesons including both nonstrange and strange ones, we 
aim at a general view on the applicability of the combination of 
HQEFT and LCSR method in  
studying heavy to light semileptonic decays. 

The paper is organized as follows: In section \ref{formulation}, we first formulate
the heavy to light transition matrix elements in the framework of HQEFT, 
and then present the light cone sum rules for the heavy flavor
independent wave functions. Some of the analytic formulae are found to be similar to 
those presented in Refs.\cite{bpi,brho} and so have a general meaning.  
In reviewing them, we pay our main attention 
to the relations and differences among different decay channels.
Section \ref{result} contains our numerical analysis on the heavy to light transition 
form factors. The numerical results are compared with data from other approaches. 
Based on the results of section \ref{result}, branching ratios are evaluated and 
discussed in section \ref{ratio}. Finally a short summary is presented 
in section \ref{sum}. 

\section{Wave functions and light cone sum rules}\label{formulation}

 For convenience of discussions, we denote in this paper
the light pseudoscalar and vector mesons as $P$ and $V$ respectively, and use
$M$ to represent the heavy mesons $B$, $B_s$ and $D$. Then the decay matrix elements
and form factors can be written in a general form as follows
\begin{eqnarray}
\label{fdefa}
&&\langle P(p)|\bar q\gamma^\mu Q|M(p+q) \rangle =2f_{+}(q^2) p^\mu+(f_{+}(q^2)+f_{-}(q^2))q^\mu,  \\
\label{fdefb}
&& \langle V(p,\epsilon^*)|\bar q\gamma^\mu (1-\gamma^5)  Q|M(p+q) \rangle =-i (m_M+m_V) A_1(q^2)
 \epsilon^{*\mu} +i \frac{A_2(q^2)}{m_M+m_V} (\epsilon^{*}\cdot (p+q) )\nonumber\\
&&\hspace{2cm} \times (2p+q)^\mu 
 +i\frac{A_3(q^2)}{m_M+m_V} (\epsilon^* \cdot (p+q)) q^\mu
   + \frac{2 V(q^2)}{m_M+m_V} \epsilon^{\mu \alpha \beta \gamma} 
    \epsilon^*_\alpha (p+q)_\beta p_\gamma, 
\end{eqnarray}
where $q$ in the currents (not to be confused with the lepton pair momentum) 
represents light quarks ($u$, $d$ or $s$),
and $Q$ denotes any heavy quark ($b$ or $c$). 
$m_M$, $m_{P(V)}$ are the heavy and light pseudoscalar (vector) meson masses, 
respectively. 
In HQEFT, the leading order matrix elements 
in the $1/m_Q$ expansion can be simply expressed as the following trace 
formulae \cite{gzmy,hly,bpi,brho}
\begin{eqnarray}
\label{parinhqet1}
\langle P(p)|\bar{q} \Gamma \QV|M_v \rangle &=&-Tr[\pi(v,p)\Gamma {\cal M}_v], \\
\label{parinhqet2}
\langle V(p,\epsilon^*)|\bar{q} \Gamma \QV|M_v \rangle &=&-i \mbox{Tr}[\Omega(v,p)\Gamma {\cal M}_v]
\end{eqnarray}
with
\begin{eqnarray}
\label{pari1}
 \pi(v,p)&=&\gamma^5 [A(v\cdot p,\mu)+ {\hat{p}\hspace{-0.2cm}\slash}
B(v\cdot p,\mu)], \\
\label{pari2}
 \Omega(v,p)&=&L_1(v\cdot p) {\epsilon\hspace{-0.2cm}\slash}^*
   +L_2( v\cdot p) (v\cdot \epsilon^*) +[L_3(v\cdot p) 
   {\epsilon\hspace{-0.2cm}\slash}^* +L_4(v\cdot p) (v\cdot \epsilon^* )]
   {\hat{p}\hspace{-0.2cm}\slash},
\end{eqnarray}
where $\hat{p}^\mu=p^\mu/v\cdot p$. $\QV$ in Eqs.(\ref{parinhqet1}), 
(\ref{parinhqet2}) is the effective heavy quark
field variable introduced in the HQEFT \cite{ylw,wwy}, which 
carries only the residual momentum $k^\mu=p^\mu-m_Q v^\mu$ with $v^\mu$ 
the heavy meson's velocity. Correspondingly $M_v$  
is the effective heavy meson state. It is related to the heavy meson 
state $M$ in Eqs.(\ref{fdefa}) and (\ref{fdefb}) by the normalization of the hadronic matrix 
elements \cite{wwy}:
\begin{eqnarray}
\label{normalization}
\frac{1}{\sqrt{m_M} }\langle \pi(\rho)|\bar q \Gamma Q|M \rangle =
\frac{1}{\sqrt{\bar\Lambda_M}}
\{  \langle \pi(\rho)|\bar{q} \Gamma \QV|M_v \rangle +O(1/m_Q) \}
\end{eqnarray}
with $\bar{\Lambda}_M = m_M - m_Q$.  
${\cal M}_v$ in Eqs.(\ref{parinhqet1}), (\ref{parinhqet2}) is the heavy pseudoscalar spin wave 
function in the HQEFT \cite{wwy}, 
\begin{equation}
{\cal M}_v = -\sqrt{\bar \Lambda}\frac{1+v\hspace{-0.2cm}\slash}{2} \gamma ^5 ,
\end{equation}
which exhibits a manifest heavy flavor symmetry.
Here $\bar{\Lambda} = \lim_{M_Q\rightarrow \infty} \bar{\Lambda}_M $ is the heavy flavor
independent binding energy.

The relations between the form factors defined in Eqs.(\ref{fdefa}), (\ref{fdefb}) and 
the universal wave functions defined in Eqs.(\ref{parinhqet1})-(\ref{pari2}) 
can be derived straightforwardly. One has 
\begin{eqnarray}
\label{AandL1}
 f_{\pm}(q^2)&=&\frac{1}{m_M} \sqrt{ \frac{m_M \bar\Lambda}
 {\bar{\Lambda}_M } }
    \{ A(v\cdot p)\pm B(v\cdot p) \frac{m_M}{v\cdot p}  \} +\cdots; \\
\label{AandL2}
A_1(q^2)&=&\frac{2}{m_M+m_V} \sqrt{\frac{m_M \bar\Lambda}{\bar\Lambda_M}} 
    \{ L_1(v\cdot p)+L_3(v\cdot p) \}  +\cdots ;  \\
\label{AandL3}
A_2(q^2)&=&2 (m_M+m_V) \sqrt{\frac{m_M \bar\Lambda}{\bar\Lambda_M}} 
    \{\frac{L_2(v\cdot p)}{2 m^2_M} +\frac{L_3(v\cdot p)
    -L_4(v\cdot p)}{2m_M (v\cdot p)} \}  +\cdots ;  \\ 
\label{AandL4}
A_3(q^2)&=&2 (m_M+m_V) \sqrt{\frac{m_M \bar\Lambda}{\bar\Lambda_M}} 
    \{ \frac{L_2(v\cdot p)}{2 m^2_M} -\frac{L_3(v\cdot p)
    -L_4(v\cdot p)}{2m_M (v\cdot p)} \}   + \cdots ; \\  
\label{AandL5}
V(q^2)&=&\sqrt{\frac{m_M \bar\Lambda}{\bar\Lambda_M}}
    \frac{m_M+m_V}{m_M (v\cdot p) } L_3(v\cdot p) +\cdots .
\end{eqnarray}

Note that the form factors introduced in Eqs.(\ref{fdefa}), (\ref{fdefb}) are heavy flavor dependent.  
But the functions $A$, $B$ and $L_i (i=1,2,3,4)$ in Eqs.(\ref{pari1}), 
(\ref{pari2}) are leading order wave functions 
in the $1/m_Q$ expansion, so they should be (at least explicitly) independent
of the heavy quark mass. 
An advantage using the effective field theory of heavy quark is that it enables 
one to formulate wave functions conveniently in such a heavy quark mass 
independent way. 
When the wave functions $A$, $B$ and $L_i (i=1,2,3) $ are estimated 
to a relatively precise extent, then the form factors $f_+$, $f_-$, $A_i(i=1,2,3)$ 
and $V$ for different decay channels can be easily obtained by adopting the relevant 
parameters such as masses and binding energies 
of the mesons. In other words, Eqs.(\ref{AandL1})-(\ref{AandL5}) show the 
relations among
different decays. From this point of view, considering the whole group of heavy to light semileptonic
decays, one can say that to certain order of the $1/m_Q$ expansion, the 
HQS and HQE simplify the theoretical analysis and reduce
the number of independent functions, though it is well known that for an 
individual decay the number of independent functions does not decrease.  

In light cone sum rule analysis, the wave functions are explored from the study of appropriate two point 
correlation functions. For example, for decays into pseudoscalar and 
vector light mesons, using the interpolating current $\bar{Q} i \gamma^5 q$ 
for the pseudoscalar heavy mesons, one may consider the functions:
\begin{eqnarray}
\label{correlator1}
P^\mu(p,q)&=&i\int d^4x e^{iq\cdot x} \langle P(p)|T\{\bar{q}(x)\gamma^\mu Q(x),
  \bar{Q}(0) i\gamma^5 q(0)\} |0 \rangle , \\
\label{correlator2}
V^\mu(p,q)&=&i\int d^4x e^{-ip_{\small M}\cdot x}  \langle V(p,\epsilon^*)|
   T\{\bar{q}(0)\gamma^\mu (1-\gamma^5)  Q(0),\bar{Q}(x)
  i\gamma^5 q(x)\} |0\rangle .
\end{eqnarray}
In applying for the sum rule method, on the phenomenological considerations,  a complete set of states with
the heavy meson quantum numbers are inserted into the above two point functions,
i.e., between the two currents. For the insertion of the ground states of heavy
mesons, one obtains meson pole contributions. While for the insertion
of higher resonances, the results are generally written in the form of integrals
over physical densities $\rho_P(v\cdot p,s)$ and $\rho_V(v\cdot p,s)$. 
In the resulting formulae, the matrix elements can be expanded in powers of
$1/m_Q$ in the effective theory. 
In this paper we consider only the leading order contributions in the heavy 
quark expansion. Similar to Refs.\cite{bpi,brho}, we have
\begin{eqnarray}
\label{phen1}
 P^\mu(p,q) & = &  2iF \frac{ A v^\mu+B \hat{p}^\mu }{2\bar\Lambda_M-2v\cdot k}
    +\int^{\infty}_{s_0} ds \frac{\rho_P(v\cdot p,s)}{s-2v\cdot k}+\mbox{subtractions},\\
\label{phen2}
 V^\mu(p,q) & = &  \frac{2F}{2\bar\Lambda_M-2v\cdot k}
   \{(L_1+L_3) \epsilon^{*\mu} -L_2 v^\mu (\epsilon^* \cdot v) 
   -(L_3-L_4) p^\mu \frac{\epsilon^* \cdot v}{v\cdot p} \nonumber\\
& - & i\frac{L_3}{v\cdot p} \epsilon^{\mu\nu\alpha\beta} \epsilon^*_\nu p_\alpha
   v_\beta \}+\int^\infty_{s_0} ds \frac{\rho_V(v\cdot p,s)}{s-2v\cdot k}
   +\mbox{subtractions} ,
\end{eqnarray}
where $F$ is the scaled decay constant of the heavy meson at the leading order of $1/m_Q$\cite{ww}.

Note that when studying $B\to \rho $ decay in \cite{brho}, there is an 
overall factor $\frac{m_B \bar{\Lambda}}{m_b \bar{\Lambda}_B}$ multiplied to 
the meson pole contribution. Though such a factor can be obtained in the 
effective field theory, it can be written in the form $1+O(1/m_Q)$. Since we 
want Eqs.(\ref{phen1}) and (\ref{phen2}) represent only the leading order contribution in heavy 
quark expansion, it is consistent to miss the factor 
$\frac{m_M \bar{\Lambda}}{m_Q \bar{\Lambda}_M}$ here. 
This change may enlarge the $B\to \rho l\nu$ decay form factors obtained in 
\cite{brho} by the same ratio, as can be seen in the next section. 

Heavy quark expansion can be performed to the correlators (\ref{correlator1}), 
(\ref{correlator2}) 
in the framework of effective theory, after which the field variables and meson 
states can be replaced by their counterparts in effective 
theory, and those correlators turns into 
\begin{eqnarray}
\label{correlatorinHQET1}
 P^\mu(p,q)&=&i\int d^4x e^{i(q-m_bv)\cdot x}
   \langle P(p)|T{\bar{q}(x)\gamma^\mu \QV(x), \QVB(0)i\gamma^5 q(0) }|0 \rangle  
 +O(1/m_Q),\\
\label{correlatorinHQET2}
 V^\mu(p,q)&=&i\int d^4x e^{-ip_M\cdot x +im_Qv\cdot x}
   \langle V(p,\epsilon^*)|T\{\bar{q}(0)\gamma^\mu (1-\gamma^5) \QV(0)
   , \QVB(x)i\gamma^5 q(x)\}|0 \rangle \nonumber\\
& +& O(1/m_Q).
\end{eqnarray}

We will include for light pseudoscalar and vector mesons the distribution 
amplitudes up to the same order as in Refs.\cite{bpi,brho}.
In other words, we consider the $\pi$ meson distribution amplitudes up to twist
4 and $\rho$ meson distribution amplitudes up to twist 2.
These $\pi$, $\rho$ distribution amplitudes are defined by
\begin{eqnarray}
\label{wfdef}
\langle \pi(p)|\bar{u}(x)\gamma^\mu \gamma^5 d(0)|0 \rangle &=&-ip^\mu f_\pi \int^1_0 du e^{iup\cdot x}
   [\phi_\pi(u)+x^2 g_1(u) ]\nonumber\\
   &+&f_\pi (x^\mu-\frac{x^2 p^\mu}{x\cdot p})
   \int^1_0 du e^{iup\cdot x} g_2(u), \nonumber\\
\langle \pi(p)|\bar{u}(x)i \gamma^5 d(0)|0 \rangle &=&\frac{f_\pi m^2_\pi}{m_u+m_d} \int^1_0 du
   e^{iup\cdot x} \phi_p(u), \nonumber\\
\langle \pi(p)|\bar{u}(x) \sigma_{\mu\nu} \gamma^5 d(0)|0 \rangle &=& i(p_\mu x_\nu-p_\nu x_\mu)
   \frac{f_\pi m^2_\pi}{6 (m_u+m_d) } \int^1_0 du e^{iup\cdot x} \phi_\sigma(u),\nonumber\\
\langle \rho(p,\epsilon^*)|\bar{u}(0) \sigma_{\mu\nu} d(x)|0 \rangle &=&-i f^\bot_\rho
  (\epsilon^*_\mu p_\nu - \epsilon^*_\nu p_\mu)
\int^1_0 du e^{iup\cdot x} \phi_\bot (u)  ,\nonumber\\
\langle \rho(p,\epsilon^*)|\bar{u}(0) \gamma_\mu d(x)|0 \rangle &=& f_\rho m_\rho p_\mu
  \frac{\epsilon^*\cdot x}{p\cdot x} \int^1_0 du e^{iup\cdot x} \phi_{\|}(u) \nonumber\\
  &+& f_\rho m_\rho (\epsilon^*_\mu-p_\mu \frac{\epsilon^* \cdot x}{p\cdot x})
  \int^1_0 du e^{iup\cdot x} g^{(v)}_\bot (u), \nonumber\\
\langle \rho(p,\epsilon^*)|\bar{u}(0) \gamma_\mu \gamma_5 d(x)|0 \rangle &=&
  \frac{1}{4} f_\rho m_\rho \epsilon_{\mu\nu\alpha\beta} \epsilon^{*\nu} p^\alpha x^\beta
  \int^1_0 du e^{iup\cdot x} g^{(a)}_\bot (u).
\end{eqnarray}
For $K$ and $K^*$ mesons, we need only replace the $\pi$ and $\rho$ mesons and $d$ quark in the
lhs. of the above equations into $K$ and $K^*$ mesons and $s$ quark, and at the same time change the
quantities and distribution amplitudes related to $\pi$ and $\rho$ mesons to those
related to $K$ and $K^*$ mesons in the rhs. of the equations. 
For decays into Kaon mesons, 
the light flavor SU(3) breaking effects may exhibit via both the light meson related
quantities and the Kaon meson distribution amplitudes. 

The standard procedure of light cone sum rule method is to calculate the correlation
functions in deep Euclidean region by using QCD or effective theories, then equate the
results with the phenomenological representations. In searching for reasonable and stable
results, the quark-hadron duality and Borel transformation are generally applied to
both sides of the equations. 
Different from the previous LCSR analysis in QCD framework \cite{pvesrb,var,ar,arsc}, 
in the current study, we instead calculate the correlation functions in HQEFT, 
i.e., we adopt the Feynman rules in effective theory. In particular, we use 
$\frac{1+\VS}{2} \int^\infty_0 dt \delta (x-y-vt)$ for the contraction of effective 
heavy quark fields $\QVB (x)$ and $\QV (y)$. 
As one can see in Refs.\cite{bpi,brho}, the theoretical calculations can 
often become simpler
in the framework of effective theory than in QCD.  
Adopting procedures similar to Refs.\cite{bpi,brho}, we get 
\begin{eqnarray}
\label{sr1}
 A(y)&=&-\frac{f_{\pi(K)}}{4F} \int^{s_0}_{0} ds e^{\frac{ 2\bar\Lambda_M-s}{T}}
    [\frac{1}{y^2} \frac{\partial}{\partial u}g_2(u)-\frac{\mu_{\pi(K)}}{y} \phi_p(u)
     -\frac{\mu_{\pi(K)}}{6y}
    \frac{\partial}{\partial u}\phi_\sigma(u) ]_{u=1-\frac{s}{2y}},\\
\label{sr2}
 B(y)&=&-\frac{f_{\pi(K)}}{4F} \int^{s_0}_{0} ds e^{\frac{ 2\bar\Lambda_M-s}{T}}
    [-\phi_{\pi(K)} (u)+\frac{1}{y^2}\frac{\partial^2}{\partial u^2}g_1(u)
    -\frac{1}{y^2} \frac{\partial}{\partial u} g_2(u) \nonumber\\
 && +\frac{\mu_{\pi(K)}}{6y}\frac{\partial}{\partial u} \phi_\sigma(u) ]_{u=1-\frac{s}{2y}}, \\
\label{sr3}
L_1(y)&=&  \frac{1}{4F} \int^{s_0}_0 ds e^{\frac{ 2\bar\Lambda_M-s}{T}}
 \frac{1}{y} f_V m_V [ g^{(v)}_\bot(u)- \frac{1}{4}
   (\frac{\partial}{\partial u} g^{(a)}_\bot (u) ) ]_{u=\frac{s}{2y}}, \\
\label{sr4}
L_2(y)&=& 0, \\
\label{sr5}
L_3(y)&=&  \frac{1}{4F} \int^{s_0}_0 ds e^{\frac{ 2\bar\Lambda_M-s}{T}}
 [\frac{1}{4y} f_V m_V (\frac{\partial}{\partial u}
   g^{(a)}_\bot(u) )+f^\bot_V \phi_\bot(u) ]_{u=\frac{s}{2y}}, \\
\label{sr6}
L_4(y)&=&  \frac{1}{4F} \int^{s_0}_0 ds e^{\frac{ 2\bar\Lambda_M-s}{T}}
 \frac{1}{y} f_V m_V [ \phi_{\|}(u)-g^{(v)}_\bot(u)
   +\frac{1}{4} (\frac{\partial}{\partial u} g^{(a)}_\bot (u) ) ]_{u=\frac{s}{2y}}.
\end{eqnarray}
$L_2(y)$ equals zero in the present approximation since no twist 2 distribution
amplitudes contribute to it. As a consequence, one can see from 
Eqs.(\ref{AandL3}), (\ref{AandL4}) that $A_2$ 
and $A_3$ have the same absolute value but opposite signs at the leading order.

 Before proceeding, we would like to address that though the formulae 
(\ref{sr1})-(\ref{sr6})
 are explicitly independent of the heavy quark mass,
it does not mean that the values of these functions for decays of
different heavy mesons are necessarily the same. This is because that  
one needs to take into account different energy scales for different heavy mesons
in calculating the corresponding functions. It is this difference that makes the distribution
amplitudes and other light meson parameters change their values 
for decays of different heavy mesons. There is also an exponent 
$e^{2\bar{\Lambda}_M/T}$ in each formula, which indicates the
dependence on the binding energy of the heavy meson.
Since $\bar{\Lambda}_B-\bar{\Lambda}_D$ is small, this exponent only introduces
a slight difference among the wave functions for $B$ and $D$ decays. 
On the other hand, there are light meson distribution amplitudes, light meson
masses and other light meson parameters in Eqs.(\ref{sr1})-(\ref{sr6}). Consequently, the resulting numerical values for the
universal wave functions may be different for different light mesons.
These heavy and light flavor symmetry breaking effects will be explored 
numerically in the following sections. 

\section{Numerical analysis of the form factors}\label{result}

The light cone distribution amplitudes embody the nonperturbative contributions, and 
they are of crutial importance for the precision that light cone sum rules can reach. 
The study of these distribution amplitudes constitutes an important and difficult project. 
They have been studied by several groups.
The asymptotic form and the scale dependence of these functions are given by perturbative 
QCD \cite{va,bf}.

For light pseudoscalar mesons, the leading twist distribution amplitude is generally 
written as an expansion in terms of the Gegenbauer polynomials $C^{3/2}_{n}(x)$ as 
follows:
\begin{eqnarray}
\label{Kfunction}
\phi_{\pi(K)}(u,\mu)&=&6u(1-u)[1+\sum^4_{n=1}a^{\pi(K)}_{n}(\mu)C^{3/2}_{n}(2u-1)].
\end{eqnarray}
One should perform the sum rule analysis at appropriate energy scale $\mu$. 
In the processes of $B_{(s)}$ and $D$ decays, the scales can be set by the 
typical virtualities of the heavy quarks, for example,   
$\mu_b=\sqrt{m^2_B-m^2_b}\approx 2.4$GeV and 
$\mu_c=\sqrt{m^2_D-m^2_c}\approx 1.3  $GeV,  
respectively \cite{vvar}. 

For $\pi$ meson, we use \cite{vvar}
\begin{eqnarray}
\label{para1}
a^{\pi}_2(\mu_c)=0.41, \;\; a^{\pi}_4(\mu_c)=0.23, \;\; a^{\pi}_1=a^{\pi}_3
=0, 
\end{eqnarray}
while for the kaon distribution amplitude $\phi_K$, when the light flavor SU(3) 
breaking effects are considered, we may use \cite{arsc} 
\begin{eqnarray}
\label{para2}
a^{K}_1 (\mu_c)=0.17, \;\; a^{K}_2(\mu_c)=0.21, \;\; a^{K}_3(\mu_c)=0.07,\;\;
a^{K}_4(\mu_c)=0.08, 
\end{eqnarray}
where the nonvanishing values of the coefficients $a_1$, $a_3$ imply the asymmetric 
momentum distributions for the $s$ and $u$, $d$ quarks inside the $K$ meson. 
Since the Gegenbauer moments $a_i$ renormalize multiplicatively, the values 
of $a_i(\mu_b)$ can be obtained from Eqs.(\ref{para1}) and (\ref{para2}) through the 
renormalization group evolution. 

We neglect the SU(3) breaking effects in the twist 3 and 4 distribution amplitudes 
included in this paper. This is justified by the analyses in Ref.\cite{pb},  
which indicates that these breaking 
effects would influence the light cone sum rules very slightly. 
Therefore we take for both $\pi$ and $K$ the following twist 3 and 4 distribution 
amplitudes \cite{ar,vvar,vi}: 
\begin{eqnarray}
\label{pifunction}
\phi_p(u)&=& 1+\frac{1}{2}B_2 [3(2u-1)^2-1]+\frac{1}{8}B_4 [35 (2u-1)^4
  -30 (2u-1)^2+3] ,\nonumber\\
\phi_\sigma(u)&=& 6u(1-u)\{ 1+\frac{3}{2}C_2 [5(2u-1)^2-1]+\frac{15}{8}C_4
  [21 (2u-1)^4-14(2u-1)^2+1]\},  \nonumber\\
g_1(u)&=& \frac{5}{2}\delta^2 u^2 (1-u)^2+\frac{1}{2}\epsilon \delta^2
  [u(1-u)(2+13u (1-u)+10u^3 \log u(2-3u+\frac{6}{5}u^2) \nonumber\\
  &+&10(1-u)^3 \log((1-u)(2-3(1-u)+\frac{6}{5}(1-u)^2)) ], \nonumber\\
g_2(u)&=&\frac{10}{3} \delta^2 u(1-u)(2u-1),
\end{eqnarray}
where 
\begin{eqnarray}
\label{para}
&&B_2(\mu_b)=0.29, \;\; B_4(\mu_b)=0.58,\;\; 
B_2(\mu_c)=0.41, \;\; B_4(\mu_c)=0.925,   \nonumber\\
&&C_2(\mu_b)=0.059, \;\; C_4(\mu_b)=0.034, \;\;
C_2(\mu_c)=0.087, \;\; C_4(\mu_c)=0.054, \;\; \nonumber\\
&&\delta^2(\mu_b)=0.17 \mbox{GeV}^2, \;\; 
\delta^2(\mu_c)=0.19   \mbox{GeV}^2, \;\; 
\epsilon(\mu_b)=0.36 ,\;\; 
\epsilon(\mu_c)=0.45   . 
\end{eqnarray}

Besides in these distribution amplitudes, the SU(3) breaking effects also 
emerge in the light meson constants in the coefficients of  
distribution amplitudes. We take 
$f_\pi=0.132 $GeV, $f_K=0.16$GeV and $\mu_\pi=m^2_{\pi}/(m_u+m_d) $ with 
$\mu_{\pi}(1\mbox{GeV})=1.65$GeV. 
For $\mu_K=m^2_K/(m_s+m_{u,d})$, we use the advocation in Ref.\cite{pb} 
to rely on chiral perturbation theory in the SU(3) limit and so 
use $\mu_K=\mu_{\pi}$.
For the quantities relevant to heavy hadrons, here we use the data evaluated 
in the previous paper Ref.\cite{ww}.
In particular, there we yielded $\bar{\Lambda}=0.53\pm 0.08$GeV,
$F=0.30\pm 0.06\mbox{GeV}^{3/2} $. 

For decays into light vector mesons, the leading twist distribution functions 
$\phi_\bot$ and $\phi_{\|}$ can also be expanded in Gegenbauer polynomials 
$C^{3/2}_n(x)$ with the coefficients running with the scale and described 
by the renormalization group method. Explicitly we have 
\begin{eqnarray}
\label{rhowfmu}
\phi_{\bot(\|)}(u,\mu)&=&6u(1-u)[ 1+\sum_{n=2,4,\cdots} a^{\bot(\|)}_n(\mu) 
   C^{3/2}_n(2u-1)  ], \nonumber\\
a^{\bot(\|)}_n(\mu)&=&a^{\bot(\|)}_n(\mu_0) (\frac{\alpha_s(\mu)}
   {\alpha_s(\mu_0)})^{(\gamma^{\bot (\|)}_n-\gamma^{\bot (\|)}_0)/(2\beta_0)},
\end{eqnarray}
where $\beta_0=11-(2/3)n_f$, and $r^{\bot}_n$, $r^{\|}_n$ are the one loop 
anomalous dimensions \cite{cleo,pdg}.
The coefficients $a^{\bot}_n$, $a^{\|}_n$ have been studied extensively 
in Ref.\cite{pvesrb}.
Here we use the values for $\rho $ and $K^*$ 
mesons presented in that paper, where the SU(3) breaking effects are 
included for $K^*$ meson. 

The functions $g^{(v)}_\bot$ and $g^{(a)}_\bot$ describe 
transverse polarizations of quarks in the longitudinally polarized 
mesons. They receive contribtuions of both twist 2 and twist 3. 
In this paper we will include only the twist 2 contributions, which 
are related to the longitudinal
distribtuion $\phi_{\|}(u,\mu)$ by Wandzura-Wilczek type relations
\cite{pvda,pvmisu}:
\begin{eqnarray}
\label{gcva}
g^{(v),twist \;\; 2}_\bot (u,\mu)&=&\frac{1}{2}[\int^u_0 dv \frac{\phi_{\|}(v,\mu)}
  {1-v} +\int^1_u dv \frac{\phi_{\|}(v,\mu)}{v} ], \nonumber\\
g^{(a),twist \;\; 2}_\bot (u,\mu)&=&2[(1-u)\int^u_0 dv \frac{\phi_{\|}(v,\mu)}
  {1-v} +u \int^1_u dv \frac{\phi_{\|}(v,\mu)}{v} ].
\end{eqnarray}

The quatities $f_\rho$ and $f_{K^*}$ are the decay constants of vector mesons, and
$f^{\bot}_{\rho}$ and $f^{\bot}_{K^*}$ are couplings defined via
\begin{eqnarray}
\langle 0|\bar{u} \sigma_{\mu\nu} q|V(p,\epsilon) \rangle =i(\epsilon_\mu p_\nu
  -\epsilon_\nu p_\nu ) f^\bot_V
\end{eqnarray}
with the light quark $q=d$ or $s$ corresponding to the vector meson 
$V=\rho$ or $K^*$. 
In the calculations, we use for these couplings \cite{pvesrb,pvda,mab,lm}:
\begin{eqnarray}
\label{paravalue}
&& f_{\rho}=195\pm 7 \mbox{MeV}, \;\;\;
f^\bot_\rho=160 \pm 10 \mbox{MeV}, \nonumber \\
&& f_{K^{*}}=226 \pm 28 \mbox{MeV}, \;\;\;\;
f^{\bot}_{K^{*}}=185 \pm 10 \mbox{MeV}.
\end{eqnarray}

 From Eqs.(\ref{AandL1})-(\ref{AandL5}) and (\ref{sr1})-(\ref{sr6}), the data for form factors can be obtained by using  the
distribution amplitudes and meson quantities presented above.
As an example, Fig.1 shows the variation of 
$A^{D\to \rho}_1$ and  $A^{D\to K^*}_1$ 
with respect to the Borel parameter $T$ at the fixed values of
$s_0=1.5$GeV, $2$GeV, $2.5$GeV. According to the light cone sum rule criterion that both the higher
resonance contributions and the contributions from higher twist distribution
amplitudes should not be too large ( say larger than 30\%), our interested 
regions of $T$ are around $T=2$GeV for $B_{(s)}$ decays, and around $T=1$GeV for 
$D$ decays. 
We studied respectively the variations of form factors in all interested decays
with respect to the Borel parameter $T$. According to our detailed study, 
for all form factors, there exist reliable regions of $s_0$ and $T$ that well satisfy
 the requirement of stability in the sum rule analysis. 
We note that in the original calculation (the first Ref. in \cite{brho}) we used a 
wrong sign for the contribution of $g^{(a)}_\bot$. 
In the latter study (the second Ref. in \cite{brho}, or an erratum 
to be published) it is found that the correction of this 
sign greatly enlarges the value of $V$, and at the same time also improves the stability 
of the $V(T)$ curves in the sum rule window. 
The threshold energies for all form factors in these decays are found to vary from 
$0.4$GeV to $3.5$GeV. 
With the ranges of $T$ and $s_0$ determined, all the 
form factors as functions of the momentum transfer $q^2$ can be derived from 
Eqs.(\ref{AandL1})-(\ref{AandL5}) and
(\ref{sr1})-(\ref{sr6}).

 It is known that for the final mesons being light pseudoscalars, the light cone
expansion and the sum rule method will break down at large momentum transfer
(numerically as $q^2$ approaches near $m^2_b$)
\cite{ar}. As a result, the curves of wave functions calculated from light cone
sum rules may become unstable at large $q^2$ region. Thus at this region we
have to rely on other approximation such as single or double
pole approximation. Here we use for $B\rightarrow \pi$ transition
\begin{eqnarray}
\label{sinpole}
f_+(q^2)=\frac{f_{B^*} g_{B^*B\pi}}{2m_{B^*}(1-q^2/m^2_{B^*})  }
\end{eqnarray}
for large $q^2$ regions of B decays into light pseudoscalar mesons.
And similar monopole approximation formulae will be applied to other heavy mesons decaying
 into light pseudoscalar mesons. In our numerical calculations, we will take:
$f_{B^*}=0.16\pm 0.03 \mbox{GeV}$, $g_{B^* B\pi}=29 \pm 3 $ \cite{ar}, 
$f_{D^*} g_{D^* D\pi}=2.7 \pm 0.8 \mbox{GeV} $, 
$f_{D^*_s} g_{D^*_s DK}=3.1 \pm 0.6 \mbox{GeV}$ \cite{arsc},
and $f_{B^*} g_{B^* B_s K}=3.88 \pm 0.31 \mbox{GeV} $ \cite{zfxt}.

 As a good approximation, for the behavior of the form factors in the whole 
kinematically accessible region, we use the parametrization
\begin{eqnarray}
\label{fitform}
F(q^2)=\frac{F(0)}{1-a_F q^2/m^2_B+b_F (q^2/m^2_B)^2},
\end{eqnarray}
where $F(q^2)$ can be any of the form factors $f_+$, $f_-$, $A_i(i=1,2,3)$ and $V$.
Thus each form factor will be parametrized by 3 parameters $F(0)$, $a_F$ 
and $b_F$ that need to be fitted.
For the form factor $f_+$, as mentioned above, since the light cone sum rules 
are most suitable for describing the low $q^2$ 
region of the form factors and the very high $q^2$ region is hard to be reached
by this approach, we shall use the light cone sum rule results at small $q^2$ region and the monopole
approximation (\ref{sinpole}) at large $q^2$ region to fit the 3 parameters
in Eq.(\ref{fitform}). 
For decays into vector mesons, we use only the light cone sum rule
predictions in fitting these parameters. This is because that 
the kinematically allowed ranges of $B$ to vector meson decays are
small compared with the ranges of $B $ to pseudoscalar meson decays, and therefore 
the sum rules are expected to yield reasonable values for most allowed regions 
of $q^2$ in the vector meson cases. 

At certain values of suitable thresholds $s_0$, we can fix the parameters for each
form factor. Our numerical results for those parameters are presented in Table 1-2. 
For the form factor $f_-$, since it is irrelevant to the decay 
rates when the lepton masses are neglected, we do not present their values  
in these tables. In this paper we calculate them directly from sum rules, 
and the results at large recoil regions will be used in the next section in 
comparing with the form factor relations derived based on the large mass and 
large energy expansion. 
 
\vspace{0.5cm}

\begin{center}
\begin{tabular}{c|c|c|c|c|c}
\hline \hline
 && \hspace{0.9cm} $F(0)$ \hspace{0.9cm} & \hspace{0.9cm} $a_F$ \hspace{0.9cm} &
\hspace{0.9cm} $b_F$ \hspace{0.9cm} & \hspace{0.5cm} $s_0$(GeV) \hspace{0.5cm} \\
\hline
$B\to \pi l\nu$ & $f_+$ & $0.35\pm 0.06$   & $1.31\pm 0.15 $
     & $0.35\pm 0.18$ & $ 2.3 \pm 0.6 $\\
\hline
$B_s\to K l\nu$ & $f_+$ & $0.47\pm 0.09$  & $1.12\pm 0.25$ & $0.34\pm 0.19$ 
     & $2.7\pm 0.8$ \\
\hline
$D\to \pi l\nu$ & $f_+$ & $0.67 \pm 0.19 $& $1.30\pm 0.30$ & $0.68\pm 0.38$ 
     & $0.8\pm 0.4$  \\
\hline
$D\to K l\nu$ & $f_+$ & $0.67\pm 0.20$ & $1.65\pm 0.43 $ & $1.28\pm 0.52$  
     & $0.8\pm 0.4$ \\
\hline \hline
\end{tabular}
\end{center}

\vspace{0cm}
\centerline{
\parbox{13cm}{
\small
\baselineskip=1.0pt
Table 1. Results for the form factor $f_+$ of heavy to light pseudoscalar 
meson decays. 
$s_0$ is the threshold at which the parameters $F(0)$, $a_F$, $b_F$ are fitted. 
} }

\vspace{0.5cm}

\begin{center}
\begin{tabular}{c|c|c|c|c|c}
\hline \hline
 && \hspace{0.9cm} $F(0)$ \hspace{0.9cm} & \hspace{0.9cm} $a_F$ \hspace{0.9cm} & 
\hspace{0.9cm} $b_F$ \hspace{0.9cm} & \hspace{0.5cm} $s_0$(GeV) \hspace{0.5cm}  \\
\hline
$B\to \rho l\nu$ & $A_1$ & $0.29 \pm 0.05$ & $0.35\pm 0.15$ & $-0.24\pm 0.12$ 
      & $2.1\pm 0.6$ \\
\cline{2-5}
   & $A_2$ & $0.28\pm 0.04$ & $1.09\pm 0.13$ & $0.20\pm 0.18$ & \\
\cline{2-5}
   & $A_3$ & $-0.28\pm 0.04$ & $1.09\pm 0.13$ & $0.20\pm 0.18$ & \\
\cline{2-5}
   & $V$ & $0.35\pm 0.06$ & $1.32\pm 0.12$ & $0.46\pm 0.10$ & \\
\hline
$B_s\to K^* l\nu$ & $A_1$ & $0.28\pm 0.05$ &$0.82\pm 0.06$ & $0.05\pm 0.14$ 
   & $2.5\pm 0.5$ \\
\cline{2-5}
   & $A_2$ & $0.28\pm 0.04$ &$1.48\pm 0.05$ & $0.62\pm 0.08$ & \\
\cline{2-5}
   & $A_3$ & $-0.28\pm 0.04$ &$ 1.48\pm 0.05$ & $0.62\pm 0.08$ & \\
\cline{2-5}
   & $V$ & $0.35\pm 0.05$ & $1.73\pm 0.02$ & $0.95\pm 0.35$ & \\
\hline
$D\to \rho l\nu$ & $A_1$ & $0.57\pm 0.08$ & $0.60\pm 0.20$ & $
0.51 \pm 1.32$ 
    & $2.0\pm 0.5$  \\
\cline{2-5}
   & $A_2$ & $0.52\pm 0.07$ &$ 0.66\pm 0.20$ & $-2.03\pm 1.52$ & \\
\cline{2-5}
   & $A_3$ & $-0.52\pm 0.07$ & $0.66\pm 0.20$ & $-2.03\pm 1.52$ & \\
\cline{2-5}
   & $V$ & $0.72\pm 0.10$ & $0.95\pm 0.06$ & $
2.60 \pm 3.62$ & \\
\hline
$D\to K^* l\nu$ & $A_1$ & $0.59\pm 0.10$ & $0.58\pm 0.12 $&$ 0.11\pm 0.28$ 
   & $2.0 \pm 0.5$\\
\cline{2-5}
   & $A_2$ & $0.55\pm 0.08$ & $0.84\pm 0.31$ & $-1.29\pm 1.12$ & \\
\cline{2-5}
   & $A_3$ & $-0.55\pm 0.08$ & $0.84 \pm 0.31$& $-1.29\pm 1.12$ & \\
\cline{2-5}
   & $V$ & $0.80\pm 0.10$ & $0.86\pm 0.65$ & $2.71\pm 1.83$ & \\
\hline \hline
\end{tabular}
\end{center}

\vspace{0cm}
\centerline{
\parbox{13cm}{
\small
\baselineskip=1.0pt
Table 2. Results for the form factors of heavy to light vector meson decays. 
$s_0$ is the threshold at which the parameters $F(0)$, $a_F$, $b_F$ are fitted. 
} }

\vspace{0.5cm}

The values of $F(0)$ for $B\to \rho$ decay presented in 
Table 2 are slightly different 
from those given in Refs.\cite{brho}. The reason is that 
the overall factor $\frac{m_b \bar{\Lambda}_B}{m_B \bar{\Lambda}}$ in 
Eq.(3.13) of \cite{brho} is missed in the current calculation, 
as mentioned in the previous section. 
Uncertainties to the form factors 
could arise from the meson constants, the light cone distribution
amplitudes and the variation of thresholds $s_0$. We notice that the latter comprises
the largest uncertainty, which may be 15\% or so. Variations of 
other imput parameters within their allowed ranges which we have discussed 
would increase the uncertainties to about 20\%. So, including uncertainties from 
higher twist amplitudes and other systematic uncertainties in light cone sum rule 
method, we quote an uncertainty of about 25\%.   

Besides the SU(3) symmetry breaking effects arise from relevant light meson 
parameters,   
it is found in our investigation that SU(3) symmetry breaking effects 
considered in the
Gegenbauer polynomial moments $a_i (i=1,2,3,4)$ invoke the changes of  
$F(0)$ by no more than 15\%. 

The form factors as functions of $q^2$ are shown in Fig.2. 
Note that in the figures some curves almost coincide with each other, 
which indicates that the SU(3) breaking effects are small in those cases. 

Table 3 is a comparison of the form factor values at $q^2=0$ obtained in this 
work and those obtained from other approaches, including 
quark models, lattice calculations, and also sum rules in QCD framework. 

\vspace{0.5cm}

\begin{center}
\begin{tabular}{c|c|c|c|c|c}
\hline \hline
  & \hspace{0.4cm} Ref. \hspace{0.4cm} &\hspace{0.7cm} $f_+(0)$\hspace{0.7cm}
  & \hspace{0.7cm} $A_1(0)$ \hspace{0.7cm}
  & \hspace{0.7cm} $A_2(0)$ \hspace{0.7cm} 
  & \hspace{0.7cm} $V(0)$ \hspace{0.7cm} \\
\hline
$ B\to \pi(\rho) l\nu $ &this work  & $ 0.35\pm 0.06 $ & $0.29\pm 0.05$ 
  & $0.28\pm 0.04 $ & $0.35\pm 0.06 $ \\
  &QM \cite{ph0001113} &0.29  & 0.26 & 0.24 & 0.31 \\
  &QM \cite{Jaus96}  & 0.27  & 0.26 & 0.24 &0.35 \\
  &Lat \cite{wuppertal} &$0.50(14)^{+7}_{-5}$   & 
       $0.16(4)^{+22}_{-16}$ & $0.72(35)^{+10}_{-7}$ 
       &  $0.61(23)^{+9}_{-6}$  \\
  &Lat \cite{lat9910021} & $ 0.28\pm 0.04 $  &  & & \\
  &SR \cite{ph9805422,ph9802394} & 0.305 & $0.26\pm 0.04$ &$0.22\pm 0.03$ 
       &$ 0.34 \pm 0.05$\\
  &SR\cite{ph0001297} &$0.28 \pm 0.05$  &  & & \\
\hline
 $B_s \to K (K^*)l\nu $ & this work & $0.47\pm 0.09 $ & $0.28 \pm 0.05$  
  & $0.28\pm 0.04 $ &$0.35 \pm 0.05 $\\
  &QM \cite{ph0001113} & 0.31 & 0.29  & 0.26 & 0.38 \\
\hline
$ D\to \pi(\rho) l\nu $ &this work  & $ 0.67\pm 0.19 $ & $0.57\pm 0.08$ 
  & $0.52\pm 0.07 $ & $0.72\pm 0.10 $ \\
  & QM \cite{ph0001113} &0.69  &0.59  &0.49 &0.90 \\
  & QM \cite{Jaus96} &0.67  &0.58  & 0.42 &0.93 \\
  & QM \cite{wsb85} & 0.69 & 0.78 & 0.92 & 1.23 \\
  &Lat \cite{wuppertal} &$0.68(13)^{+10}_{-7}$  &$0.59(7)^{+8}_{-6}$  
       & $0.83(20)^{+12}_{-8}$  & $1.31(25)^{+18}_{-13}$ \\
  &Lat \cite{lat9910021} & $0.64(5)^{+00}_{-07}$ &  & & \\
  &Lat \cite{lat9710057}  &$0.65 \pm 0.10 $ & $ 0.65 \pm 0.07 $ & $0.55\pm 0.10$ 
        & $ 1.1\pm 0.2$ \\
  &SR \cite{ph0001297} & $0.65 \pm 0.11$ &  & & \\
  &SR \cite{sr9193} &$0.50 \pm 0.15 $ &$0.5 \pm 0.2 $ &$0.4 \pm 0.2 $& $1.0 \pm 0.2 $ \\
\hline
$ D\to K(K^*) l\nu $ &this work  & $ 0.67\pm 0.20 $ & $0.59\pm 0.10$ 
  & $0.55\pm 0.08 $ & $0.80\pm 0.10 $ \\
  &E791 \cite{e791}  &   & $0.58\pm 0.03 $ & $0.41\pm 0.06$ 
      & $ 1.06\pm 0.09 $\\
  & QM \cite{ph0001113} & 0.78  &0.66  &0.49 & 1.03 \\
  & QM \cite{Jaus96} & 0.78  &0.66  &0.43 &1.04 \\
  &Lat  \cite{wuppertal} & $0.71(12)^{+10}_{-7}$   & $0.61(6)^{+9}_{-7}$ 
        &$ 0.83(20)^{+12}_{-8}$ & $1.34(24)^{+19}_{-14}$ \\
  & Lat \cite{lat9710057}& $0.73\pm 0.07$  & $0.70\pm 0.07 $ & $ 0.6\pm 0.1$ 
         & $ 1.2 \pm 0.2 $ \\
  &SR \cite{ph0001297} & $0.76\pm 0.03$  &  & & \\
  & SR \cite{sr9193}  & $0.60\pm 0.15$ & $0.50 \pm 0.15 $ & $ 0.60\pm 0.15 $ 
          & $ 1.10 \pm 0.25$ \\
\hline \hline
\end{tabular}
\end{center}

\vspace{0cm}
\centerline{
\parbox{13cm}{
\small
\baselineskip=1.0pt
Table 3. Comparison of the results for semileptonic decay form factors 
(at $q^2=0$) from different approaches (QM: quark model; Lat: lattice 
calculation; SR: sum rules in QCD framework; E791: data extracted from 
experimental measurements). 
} }

\vspace{0.5cm}

The form factor ratios $R_V \equiv V(0)/A_1(0)$, $R_2 \equiv 
A_2(0)/A_1(0)$ for $D \to K^* l\nu$ decay have been measured by several groups. 
We present in Table 4 the comparison of our results for these ratios with the  
measurements. 
Our result for $R_V$ agrees well with the latest measurement \cite{beatrice} 
but smaller than other measurements. $R_2$ obtained in this work is also in 
good agreement with the BEATRICE measurement \cite{beatrice}. 

\vspace{0.5cm}

\begin{center}
\begin{tabular}{c|c|c|c}
\hline \hline
 & \hspace{1.4cm} $R_V$ \hspace{1.4cm} & \hspace{1.4cm} $R_2$ \hspace{1.4cm} & 
\hspace{0.9cm} $A_1(0)$ \hspace{0.9cm}  \\
\hline
   this work & $1.36 \pm 0.39$ & $ 0.93 \pm 0.29 $ & $ 0.59\pm 0.10$ \\
\hline
   BEATRICE \cite{beatrice} & $1.45 \pm 0.23\pm 0.07$ & $1.00\pm 0.15 \pm 0.03$ &   \\
\hline
   E791 \cite{e791} & $1.84 \pm 0.11 \pm 0.08 $ & $0.71 \pm 0.08 \pm 0.09$ 
     & $0.58 \pm 0.03$  \\
\hline
   E687 \cite{e687} & $1.74\pm 0.27\pm 0.28 $ & $0.78 \pm 0.18\pm 0.10$ &   \\
\hline
   E653 \cite{e653} & $2.00^{+0.34}_{-0.32} \pm 0.16$ & $0.82^{+0.22}_{-0.23} \pm 0.11 $ &   \\
\hline
   E691 \cite{e691} & $2.0 \pm 0.6\pm 0.3 $ & $0.0 \pm 0.5 \pm 0.2$ & \\
\hline \hline
\end{tabular}
\end{center}

\vspace{0cm}
\centerline{
\parbox{13cm}{
\small
\baselineskip=1.0pt
Table 4. Comparison of measurements and theoretical predictions for form 
factor ratios and $A_1(0)$. 
} }

\vspace{0.5cm}

With the consideration that in heavy to light decays the final light meson 
usually carries a large recoil momentum, the large energy effective theory 
(LEET) \cite{dugan,uag,jalo,mth,ph0107065} has been proposed to study heavy to light transitions in 
the region of large energy release. In that framework and neglecting QCD short 
distance corrections, it is quoted that there are only three form factors 
needed describing all the pseudoscalar to pseudoscalar or vector transition 
matrix elements in the leading order of heavy quark mass and large energy 
expansion. 
In particular, taking into acount contributions of second order in the 
ratio of the light meson mass to the large recoil energy, the recent 
work \cite{ph0107065} derived interesting relations concerning form factors 
for semileptonic B 
decays to light pseudoscalar and vector mesons: 
\begin{eqnarray}
\label{leetebert1}
&&f_+(q^2)=\frac{m_M}{2E_F}(1+\frac{m^2_P}{m^2_M})f_0(q^2), \\
\label{leetebert2}
&&\frac{m_M}{m_M+m_V} \frac{\sqrt{E^2_F-m^2_V}}{E_F} V(q^2) =\frac{m_M+m_V}{2E_F}A_1(q^2) 
    \nonumber\\
&& \hspace{5cm} =\frac{m_M}{m_M+m_V}(1-\frac{2m^2_V}{m^2_M})A_2(q^2)+\frac{m_V}{E_F}A_0(q^2),
\end{eqnarray}
where $E_F$ is the on-shell energy of the final light meson: 
\begin{eqnarray}
E_F=\frac{m^2_M+m^2_{P(V)}-q^2}{2m_M}.
\end{eqnarray}

$f_0(q^2)$ and $A_0(q^2)$ are form factors defined in the way of 
Ref.\cite{ph0107065}, 
and they can be represented by the form factors defined in this paper as follows,
\begin{eqnarray}
f_0(q^2)&=&\frac{q^2}{m^2_M-m^2_P} f_-(q^2)+f_+(q^2), \\
A_0(q^2)&=&\frac{1}{2m_V}[(m_M+m_V)A_1(q^2)-(m_M-m_V)A_2(q^2)
     -\frac{q^2}{m_M+m_V}A_3(q^2)].
\end{eqnarray} 
As a result, relations (\ref{leetebert1})-(\ref{leetebert2}) are equivalent to 
\begin{eqnarray}
\label{leet1}
f_+(q^2)&=&\frac{m^2_M+m^2_P}{m^2_M+m^2_P-q^2} f_0(q^2), \\
\label{leet2}
V(q^2)&=&\frac{(m_M+m_V)^2}{2m_M \sqrt{E^2_F-m^2_V}} A_1(q^2),\\
\label{leet3}
A_2(q^2)&=&\frac{m^2_M q^2}{2m^2_V q^2 -2 m^4_V -m^2_M q^2} A_3(q^2).
\end{eqnarray}
Note that in the heavy quark and large recoil expansion, the leading order 
contribution of the denominator in Eq.(\ref{leet3}) is just $- m^2_M q^2 A_2(q^2)$. 
As a result, one gets $A_2(q^2)=-A_2(q^2)$ at the leading order of LEET, 
which is consistent with our calculation in this work (see section 
\ref{formulation}). 

One can now have a numerical comparison between our results for form factors with 
the LEET predictions.  
In Figs.3-5, we show the form factor ratios $f_+(q^2)/f_0(q^2)$, $V(q^2)/A_1(q^2)$ 
and $A_2(q^2)/A_3(q^2)$ derived from our sum rule calculations 
and from the relations (\ref{leet1})-(\ref{leet3}). 
In Fig.3, the ratio $f_+(q^2)/f_0(q^2)$ from our calculation and that from  
the LEET relation (\ref{leet1}) are close to each other at regions near 
the maximum recoil point, but they quickly split from each other as $q^2$ 
increases. Similar variations can be observed for the $B_{(s)}$ decay 
ratio $V(q^2)/A_1(q^2)$ in Fig.4(a)-(b). For $D$ decays to light vector mesons, 
however, in Fig.4(c)-(d) large difference exists between our results and 
the predictions of currently known LEET relation (\ref{leet2}). 
In Fig.5(a)-(b) (i.e., for $B_{(s)}$ decays), the ratio $A_2(q^2)/A_3(q^2)$ 
of our results and the LEET prediction are compatible in almost the whole 
kinematically allowed regions except the region very near to $q^2=0$ point. 
In Fig.5(c)-(d), our results for the $D$ decay ratio $A_2(q^2)/A_3(q^3)$  
and the LEET relation (\ref{leet3}) are again incompatible. 

As a general view, our results are compatible with the relations 
(\ref{leet1})-(\ref{leet3}) at appropriate regions for $B_{(s)}$ decays. 
On the other hand, to the expansion order considered here, discrepancy 
occurs between the two methods in studying $D$ decays. 
One reason for this may be of course that the charm quark (hadron) 
mass is not very large, and therefore large $1/m_c$ corrections both to our 
results of leading order in HQE and to the LEET approximations 
(\ref{leet1})-(\ref{leet3}) are possible. 
This should be clarified by the future study. 
 
\section{Branching ratios}\label{ratio}

With the form factors derived in the previous sections, 
one can extract the values of relevant CKM matrix elements from 
the experimental measurements of branching ratios. 
In Refs.\cite{bpi} and \cite{brho}, $|V_{ub}|$ is extracted in this 
way from $B \to \pi(\rho)l\nu$ decays 
with using the branching ratio measurements. The results obtained there were
$|V_{ub}|=( 3.4 \pm 0.5 \pm 0.5 )\times 10^{-3} $
and 
$|V_{ub}|=(3.7\pm 0.6 \pm 0.7 )\times 10^{-3}$ 
via the two decays, respectively, where the first (second) errors correspond 
to the experimental (theoretical) uncertainties. 

On the other hand, the decay widths and 
branching ratios of exclusive semileptonic decays can be predicted 
if we know the values of relevant CKM matrix elements. 
When the lepton masses are neglected, we have for decays to pseudoscalar 
mesons:
 \begin{eqnarray}
\label{pigammaq2}
\frac{d\Gamma}{dq^2}=\frac{G^2_F |V_{qQ}|^2}{24 \pi^3} (
(\frac{m^2_M+m^2_P-q^2}{2m_M})^2 -m^2_P)^{3/2}
   [f_+(q^2)]^2  ,
\end{eqnarray}
and for decays to vector mesons: 
\begin{eqnarray}
\label{rhogammaq2}
\frac{d\Gamma}{dq^2}=\frac{G^2_F |V_{qQ}|^2}{192 \pi^3 m^3_M} 
  \lambda^{1/2} q^2 (H^2_0+H^2_+ + H^2_-) 
\end{eqnarray}
with $\lambda \equiv (m^2_M+m^2_V-q^2)^2-4m^2_M m^2_V$ and 
the helicity amplitudes defined as  
\begin{eqnarray}
\label{heliampli}
H_{\pm}&=&(m_M+m_V) A_1(q^2) \mp \frac{\lambda^{1/2}}{m_M+m_V} V(q^2),\nonumber\\
H_0&=&\frac{1}{2m_V \sqrt{q^2}} \{ (m^2_M-m^2_V-q^2) (m_M+m_V) A_1(q^2) 
    -\frac{\lambda}{m_M+m_V}A_2(q^2)  \}.
\end{eqnarray} 

In this paper we would like to predict the branching ratios from the values 
of CKM matrix elements: $|V_{ub}|=0.0037$, $|V_{cd}|=0.22$, $|V_{cs}|=0.97$. 
Finishing the integration over $q^2$, we obtain the widths and branching ratios 
in Table 5. In that table we also list the experimental measurements of 
branching ratios given by Ref. \cite{epjc}.

\vspace{0.5cm}

\begin{center}
\begin{tabular}{c|c|c|c}
\hline \hline
 & \hspace{0.3cm} $\Gamma$ (in $|V_{qQ}|^2 \mbox{ps}^{-1}$) \hspace{0.3cm} & 
\hspace{1.8cm} Br \hspace{1.8cm} & \hspace{0.4cm} Br(measurements) \hspace{0.4cm}  \\
\hline
$B\to \pi l\nu$ & $10.2 \pm 1.5$ & $ (2.1\pm 0.5) \times 10^{-4}$ 
     & $(1.8\pm 0.6) \times 10^{-4}$ \\
\hline
$B_s\to K l\nu$ & $14.3\pm 2.8 $ & $(2.9\pm 0.7) \times 10^{-4} $ & \\
\hline
$B\to \rho l\nu$ & $15.2 \pm 4.2$ & $ (3.2 \pm 1.0) \times 10^{-4}$ 
     & $ (2.6^{+0.6}_{-0.7} ) \times 10^{-4} $ \\
\hline 
$B_s \to K^* l\nu $ & $20.2\pm 4.3$ & $(4.1\pm 1.0) \times 10^{-4}$ &\\
\hline
$D\to \pi l\nu$ &$ 0.152 \pm 0.042$ & $ (3.0\pm 0.9) \times 10^{-3}$ 
     & $(3.7\pm 0.6)\times 10^{-3}$  \\
\hline
$D\to K l\nu$ & $0.101 \pm 0.030$ & $(3.9 \pm 1.2) \times 10^{-2}$ 
     & $(3.47\pm 0.17)\times 10^{-2}$  \\
\hline
$D\to \rho l\nu$ & $0.071\pm 0.015$ & $ (1.4 \pm 0.3) \times 10^{-3}$ &  \\
\hline
$D\to K^* l\nu$ & $0.050 \pm 0.014$ & $( 2.0 \pm 0.5) \times 10^{-2} $ 
     & $ (2.02\pm 0.33) \times 10^{-2} $ \\
\hline \hline
\end{tabular}
\end{center}

\vspace{0cm}
\centerline{
\parbox{13cm}{
\small
\baselineskip=1.0pt
Table 5. Decay widths $\Gamma$ (in unit $|V_{qQ}|^2 \mbox{ps}^{-1}$) and 
branching ratios (Br) for 
heavy to light meson decays. In deriving the branching ratios we used 
$|V_{ub}|=0.0037$, $|V_{cd}|=0.22$, $|V_{cs}|=0.97$ and the lifetimes 
of heavy mesons: 
$\tau_{B^0}=1.56\pm 0.06\mbox{ps}$, $\tau_{D^0}=0.4126\pm 0.0028 \mbox{ps}$, 
$\tau_{B_S}=1.493\pm 0.062\mbox{ps}$. 
} }

\vspace{0.5cm}
The largest uncertainty for the branching ratios in Table 5 is about 
30\%.   
So, the precision of heavy meson decay measurements expected in the 
new B factories can not be matched well unless these theoretical 
uncertainties could be reduced.  
This reduction may be reached by consideration of both higher twist 
contributions, better determination of the meson constants and 
the higher order contributions in the heavy quark expansion, which should 
be discussed in the future work. 

Table 6 is a comparison of the results for the ratios $\Gamma_L/\Gamma_T$, 
$\Gamma_+/\Gamma_-$ 
and the total decay rates $\Gamma$ for 
heavy to light vector decays, 
where $\Gamma_L (\Gamma_T)$ and $\Gamma_+ (/\Gamma_-)$ represent partial widths 
for longitudinal (transverse) polarization and positive (negative) helicity, 
respectively. These widths and ratios have been predicted via other approaches. 
In particular, the ratios for  
$D\to K^* l\nu$ decay are available from recent experiments. 

\begin{center}
\begin{tabular}{c|c|c|c|c}
\hline \hline
  & \hspace{1.1cm} $\Gamma_L/{\Gamma_T}$ \hspace{1.1cm} 
  & \hspace{1.1cm} $\Gamma_+/{\Gamma_-}$ \hspace{1.1cm} &
  \hspace{0.2cm} $\Gamma/{|V_{qQ}|^2} (\mbox{ps}^{-1})  $ \hspace{0.2cm} & Ref. \\
\hline
 $B\to \rho l\nu$ & $ 0.85\pm 0.09 $  
    &$0.04 \pm 0.02$ & $15.2\pm 4.2$ & this work\\
  & 0.82 &  & $19.1  $  & QM \cite{Jaus96} \\
  & $0.88 \pm 0.08$ &  & $15.8 \pm 2.3 $ 
       & QM \cite{ph9807223} \\
  &  &  & $13\pm 12$  & Lat\cite{elc}  \\
  & $0.80^{+0.04}_{-0.03}$ &  & $16.5^{+3.5}_{-2.3} $ 
     & Lat \cite{lat98} \\
  & $0.52\pm 0.08$ &  & $13.5 \pm 4.0 $  
      & SR \cite{ph9701238} \\
  & $0.06 \pm 0.02 $ &$ 0.007\pm 0.004$  & $ 12\pm 4 $ 
        & SR \cite{sr9193} \\
\hline
 $B_s\to K^* l\nu$ & $0.79 \pm 0.08$  & $0.07 \pm 0.02$ &$20.2 \pm 4.3$ 
    &this work \\
\hline 
 $D\to \rho l\nu$ & $1.17 \pm 0.09$ & $0.29 \pm 0.13$ &$0.071 \pm 0.015$&this work \\
  & 1.16 & & $ 0.087 $ & QM \cite{ph0001113} \\
  & 0.67 &  & $0.025 $  & QM \cite{isgw2} \\
  &  &  & $12.17\pm 4.06$  & Lat \cite{ape} \\
  &  &  & $10.15 \pm 4.67$  & Lat \cite{elc} \\
  & $1.31\pm 0.11$ & $0.24\pm 0.03 $ & $0.024 \pm 0.007
       $ & SR \cite{sr9193} \\
\hline 
$D\to K^* l\nu$  & $1.15 \pm 0.10 $ & $0.32 \pm 0.13$ &$ 0.050 \pm 0.014$ 
       & this work\\
  & $1.09 \pm 0.10 \pm 0.02$& $0.28\pm 0.05\pm 0.02$ & & BEATRICE \cite{beatrice} \\
  & $1.20 \pm 0.13 \pm 0.13$ &  & & E687 \cite{e687} \\
  & $1.18 \pm 0.18 \pm 0.08$ & & &E653 \cite{e653} \\
  & $1.8^{+0.6}_{-0.4} \pm 0.3$ &  &  & E691 \cite{e691} \\
  & $0.6 \pm 0.3^{+0.3}_{-0.1} $ &  & &  WA82 \cite{wa82} \\
  & 1.28 &  & $0.063 $ & QM \cite{ph0001113}  \\
  & 1.33 &  &$ 0.058 $   & QM  \cite{Jaus96} \\
  &$1.2\pm 0.3$  &  & $0.073\pm 0.019 $  & Lat \cite{ape} \\
  &$1.1\pm 0.2$  &  & $6.0^{+0.8}_{-1.6}$  &Lat \cite{ukqcd} \\
  & $0.86 \pm 0.06$ &$3.8 \pm 1.5$  &   & SR \cite{sr9193} \\
\hline \hline
\end{tabular}
\end{center}

\vspace{0cm}
\centerline{
\parbox{13cm}{
\small
\baselineskip=1.0pt
Table 6. Theoretical predictions and measurements of the ratios  
$\Gamma_L/{\Gamma_T}$, $\Gamma_+/{\Gamma_-}$ and the decay 
rates $\Gamma$ (in unit $|V_{qQ}|^2 \mbox{ps}^{-1}$). The data from 
BEATRICE, E687, E653, E691 and WA82 are experimental measurements. 
} }

\vspace{0.5cm}

Similarly, Table 7 presents a comparison of results for heavy to light 
pseudoscalar decay rates. 

\vspace{0.5cm}

\begin{center}
\begin{tabular}{c|c|c|c|c}
\hline \hline
 Ref.  & \hspace{0.5cm} $B\to \pi l\nu $ \hspace{0.5cm} 
   & \hspace{0.5cm} $ B_s \to Kl\nu$ \hspace{0.5cm} 
  & \hspace{0.5cm} $D \to \pi l\nu$ \hspace{0.5cm} 
  & \hspace{0.5cm} $D\to Kl\nu$ \hspace{0.5cm} \\
\hline
 this work & $10.2\pm 1.5$ & $14.3 \pm 2.8$ & $0.152 \pm 0.042$ & 
   $0.101 \pm 0.030 $\\
 QM \cite{ph0001113} &   &  & $ 0.196 $ & $0.102 $ \\
  QM \cite{Jaus96} & $10.0 $  &  
       &$0.165 $ & $ 0.101 $ \\ 
  Lat \cite{ape} & $8\pm 4$ &  & $0.162\pm 0.041$  &$0.096\pm 0.021$ \\ 
  Lat \cite{elc} & $12\pm 8$  & & $0.114\pm 0.073$ & $0.072\pm 0.036$ \\ 
  Lat \cite{lat98} & $8.5^{+3.3}_{-1.4} $  &  &  & \\ 
 SR \cite{ph0001297} & $7.3\pm 2.5  $  &  
       &$0.13\pm 0.05 $  & $0.094\pm 0.036  
 $ 
 \\ 
 SR \cite{sr9193} & $5.1\pm 1.1 $  &  
       & $0.080 \pm 0.017  $   & $ 0.068\pm 0.014 $ \\ 
\hline \hline
\end{tabular}
\end{center}

\vspace{0cm}
\centerline{
\parbox{13cm}{
\small
\baselineskip=1.0pt
Table 7. Rates in unit $|V_{qQ}|^2 \mbox{ps}^{-1}$ for semileptonic decays to pseudoscalar 
mesons from different approaches.
} }

\section{Summary}\label{sum}

 In summary, we have studied the exclusive semileptonic decays of heavy to light mesons. 
The form factors for the decays $B(D, B_s) \to
\pi(\rho, K, K^*)l\nu$ have been consistently calculated by using the light cone sum 
rule method in the effective field theory of heavy quark. The HQS 
leads to simplification in studying heavy to light tansitions as to certain order 
of the $1/m_Q$ expansion the decays of different heavy hadrons such as $B$ and
$D$ can be characterized by the same set of wave functions which are explicitly 
independent of the heavy quark mass, though the HQS does not reduce the number of 
independent form factors needed for an individual decay.   
In such calculations, the uncertainties for the form factors are generally
about 25\%, which, together with the meson constants, may give the branching ratios with
a total uncertainty up to 30\%.
We have also estimated the light flavor 
SU(3) symmetry breaking effects in these semileptonic decays and found that
those effects may influence the form
factors up to a total amount of 15\%. 
Our results have been compared with data from experiments and from other theoretical 
approaches. In particular, we have checked the compatibility between our results 
and the form factor relations derived in 
Ref.\cite{ph0107065} using the heavy quark and large recoil expansion. 
We conclude that the form factors and branching ratios of those heavy to light meson 
semileptonic decays can be calculated consistently based on the light cone 
sum rule approach within the framework of HQEFT. 
Nevertheless, in order to match the expected more precise experimental 
measurements at B factories in the near 
future, more accurate calculations for the exclusive heavy to light meson 
semileptonic decays are needed. 
The uncertainties in our results and the discrepancy existing 
between our results and the relations (\ref{leet1})-(\ref{leet3}) indicate 
that the $1/m_Q$ corrections may be important in charm meson decays.  
This should be investigated in the future work. 

In this paper we have only considered the semileptonic heavy to light decays, 
but what we used here is a general approach, and similar procedures could be applied 
straightforwardly to heavy to light rare decays such as $B\to Kll$, 
$B\to K^* ll$ and $B\to K^* \gamma$. For such rare decays, this framework 
becomes more interesting because it reduces the number of form factors 
for an individual decay. We will discuss that in another paper. 
 
\acknowledgments

This work was supported in part by the key projects of
National Science Foundation of China (NSFC)  and Chinese Academy of Sciences.

\newpage
\centerline{\large{FIGURES}}

\newcommand{\PICLI}[2]
{
\begin{center}
\begin{picture}(500,120)(0,0)
\put(0,-20){
\epsfxsize=7cm
\epsfysize=10cm
\epsffile{#1} }
\put(115,40){\makebox(0,0){#2}}
\end{picture}
\end{center}
}

\newcommand{\PICRI}[2]
{
\begin{center}
\begin{picture}(300,0)(0,0)
\put(160,10){
\epsfxsize=7cm
\epsfysize=10cm
\epsffile{#1} }
\put(275,69){\makebox(0,0){#2}}
\end{picture}
\end{center}
}


\small
\mbox{}
{\vspace{1.2cm}}

\PICLI{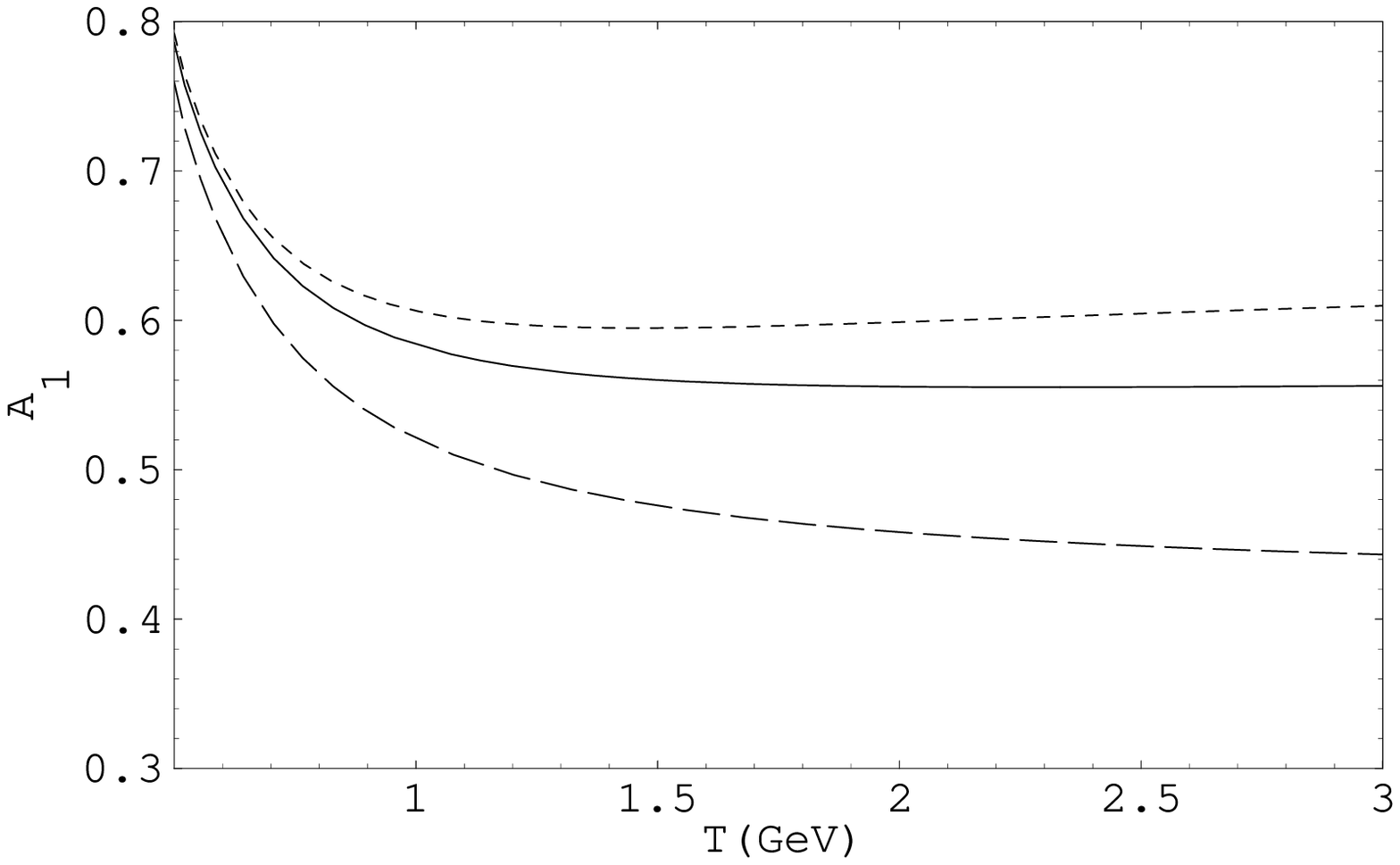}{(a)}

\PICRI{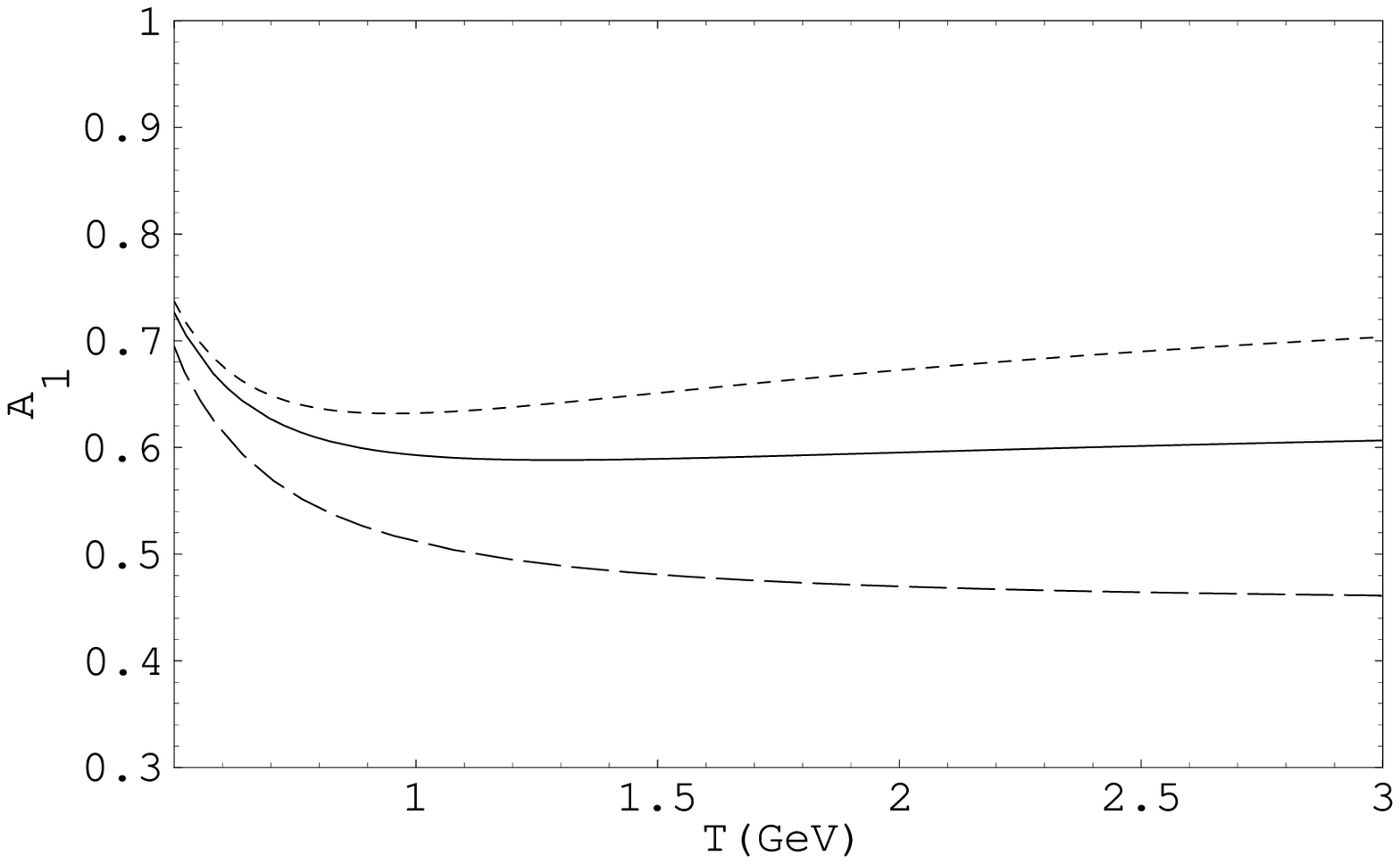}{(b)}

\vspace{-2cm}
\centerline{
\parbox{12cm}{
\small
\baselineskip=1.0pt
Fig.1. The form factors $A^{D\to \rho}_1$ ((a)) and $A^{D\to K^*}_1$ ((b)) as functions of 
the Borel parameter $T$ for different values of the continuum threshold $s_0$. 
The dashed, solid and dotted curves correspond to $s_0=$1.5, 2 and 2.5 GeV respectively. 
Considered here is at the momentum transfer $q^2=0\mbox{GeV}^2$.}}

\newpage

\mbox{}
{\vspace{1.2cm}}

\PICLI{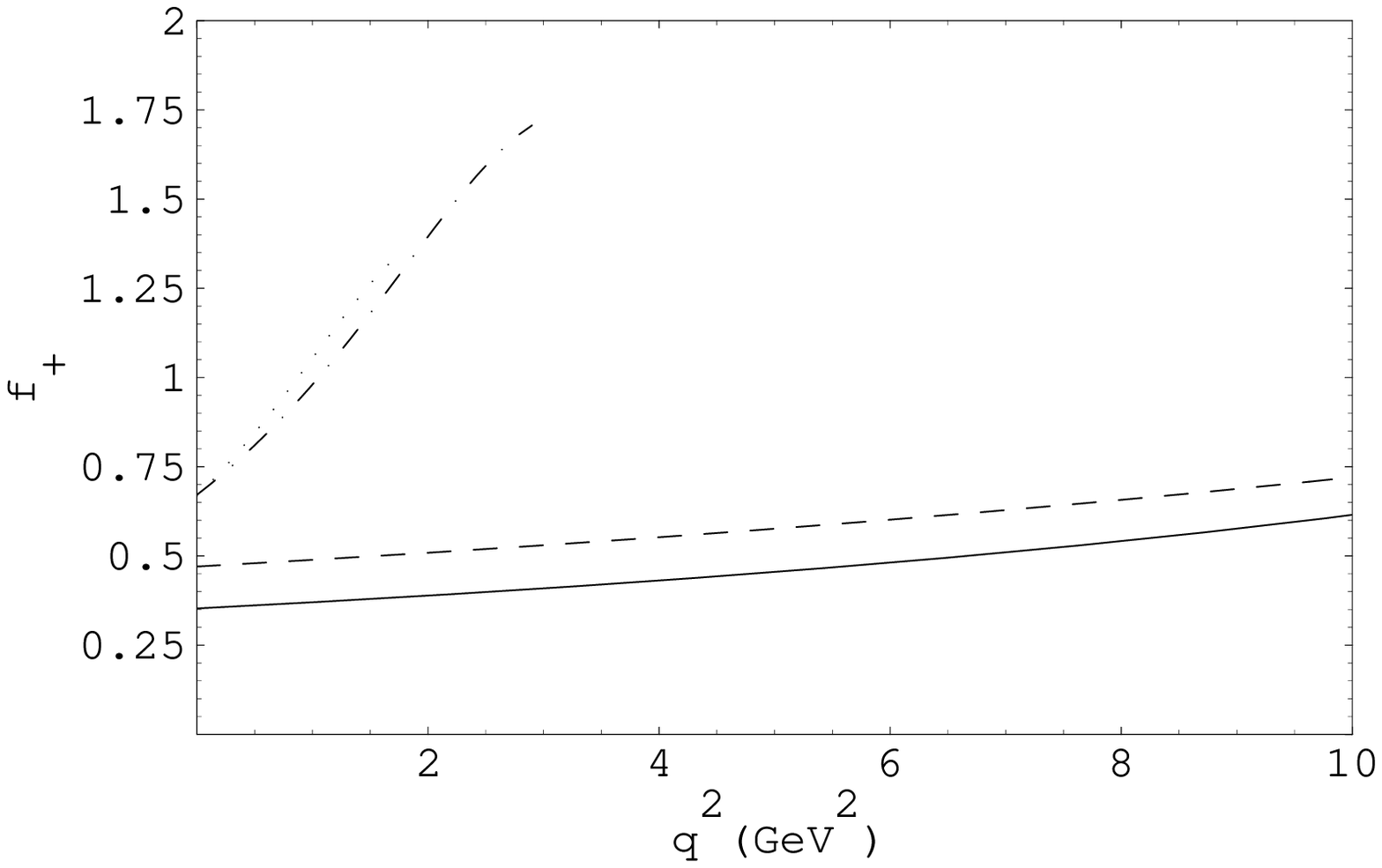}{(a)}

\PICRI{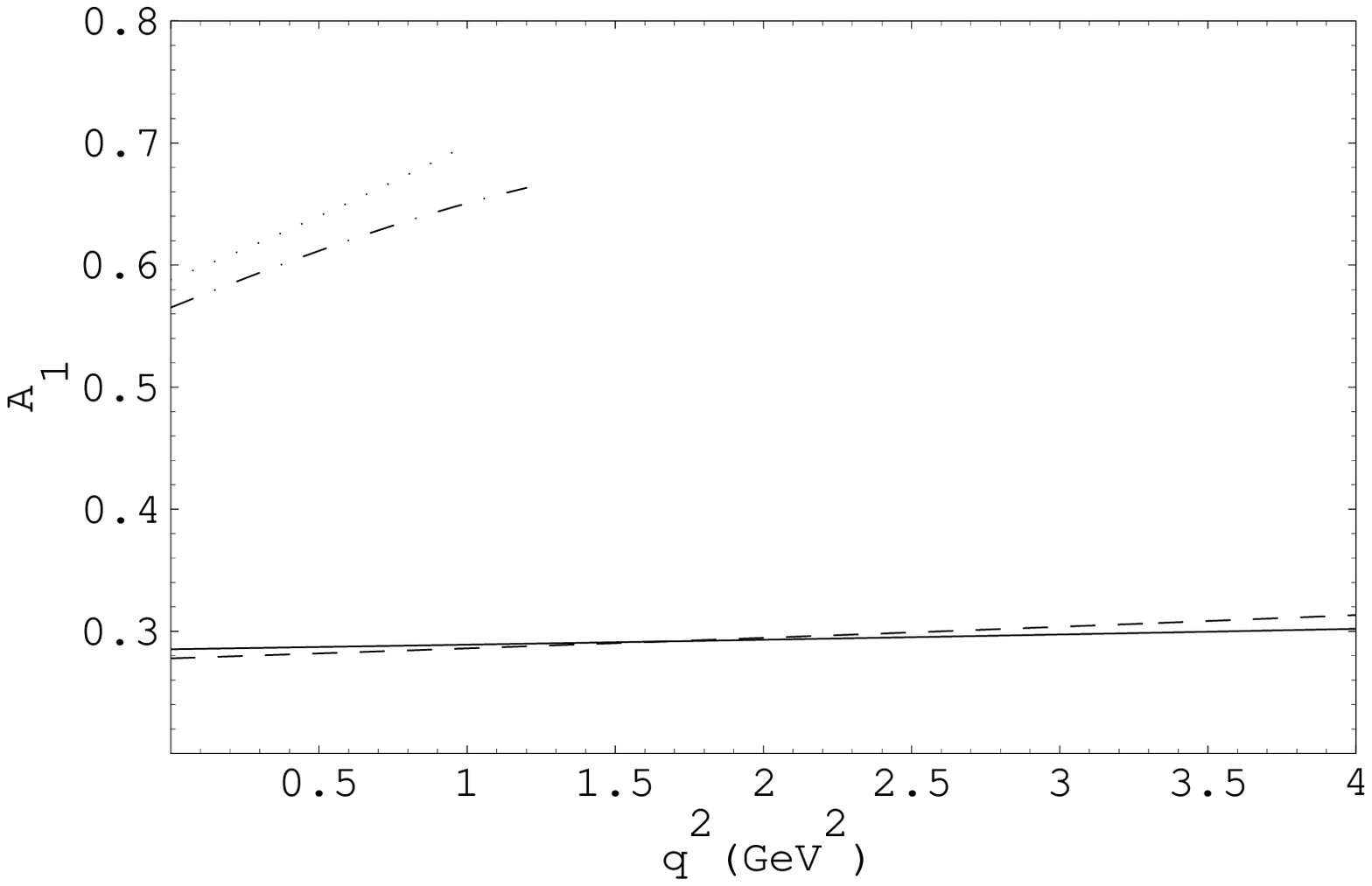}{(b)}

\PICLI{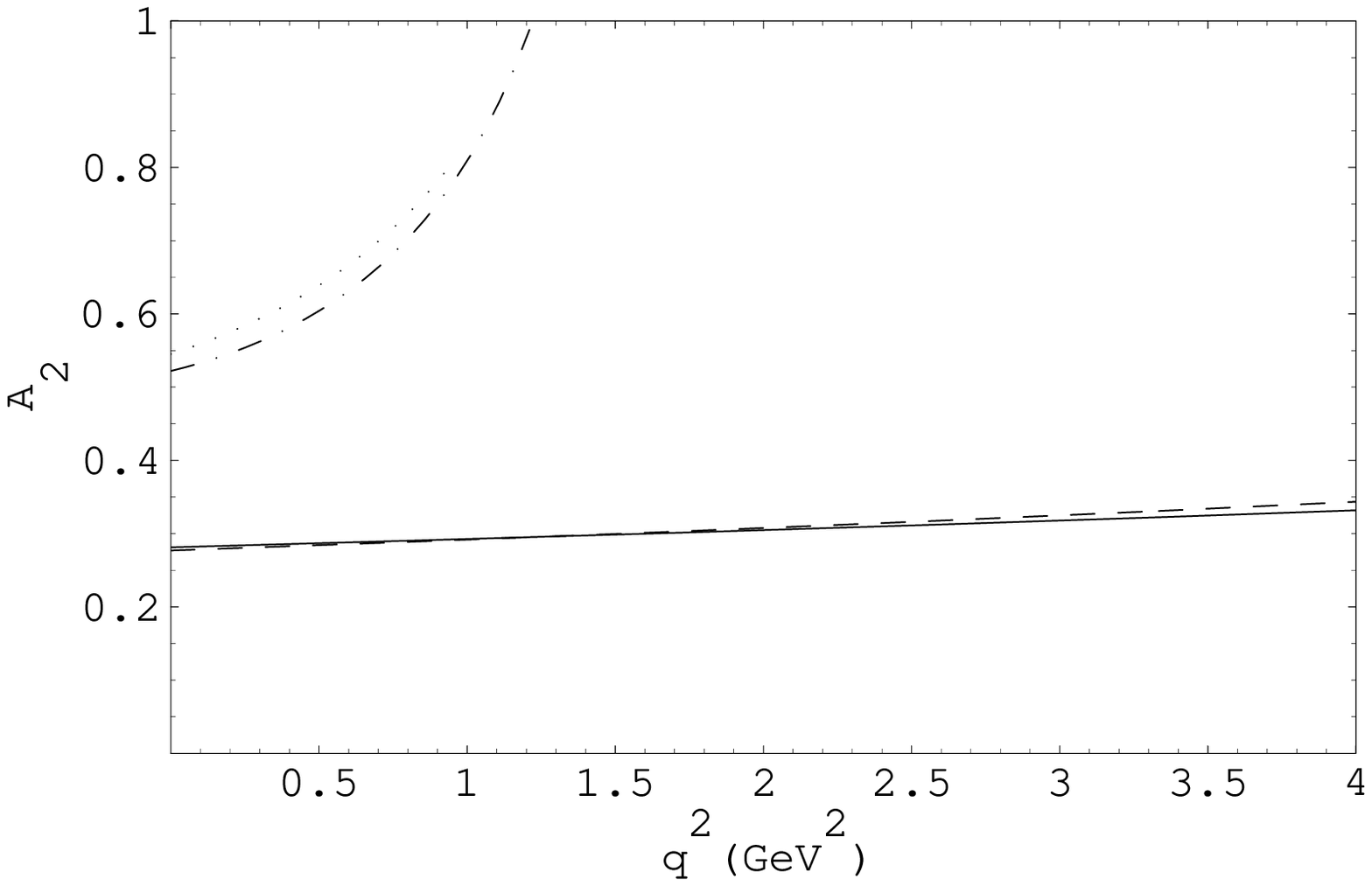}{(c)}

\PICRI{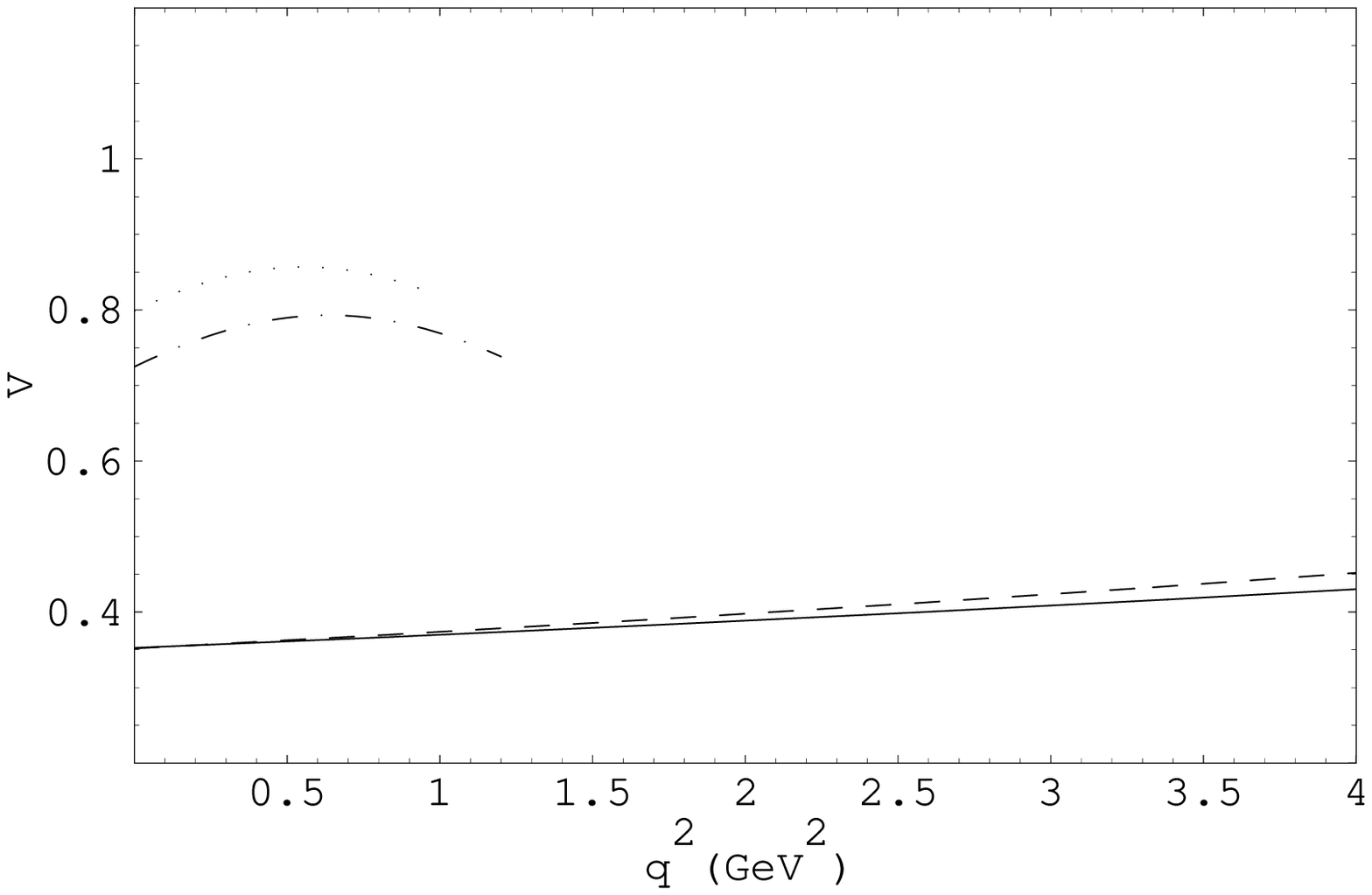}{(e)}

\vspace{-2cm}
\centerline{
\parbox{12cm}{
\small
\baselineskip=1.0pt
Fig.2. Results for the heavy-to-light decay form factors from light cone sum 
rule study. The solid, dashed, dot-dashed and dotted curves correspond to 
$B\to \pi(\rho)$, $B_s \to K(K^*)$, $D\to \pi(\rho)$ and $D\to K(K^*)$ decays, 
respectively. The dot-dashed and dotted curves in figure (a) almost coincide 
with each other, and so do the solid and dashed curves in figures (b)-(d). 
$A_3(q^2)$ is not shown here as one has $A_3(q^2)=-A_2(q^2)$ at the leading 
order considered. 
} }

\newpage

\mbox{}
{\vspace{1.2cm}}

\PICLI{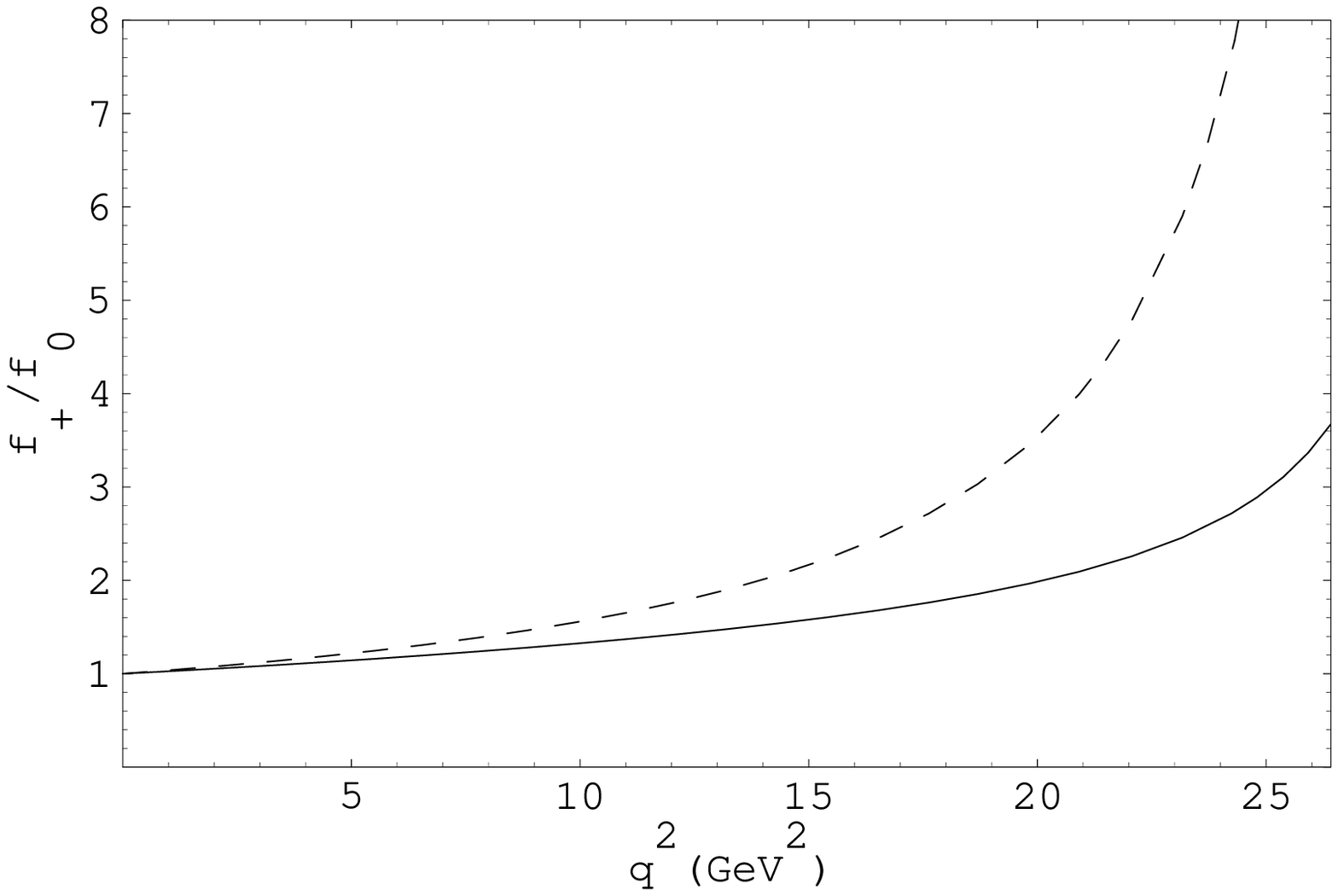}{(a)}

\PICRI{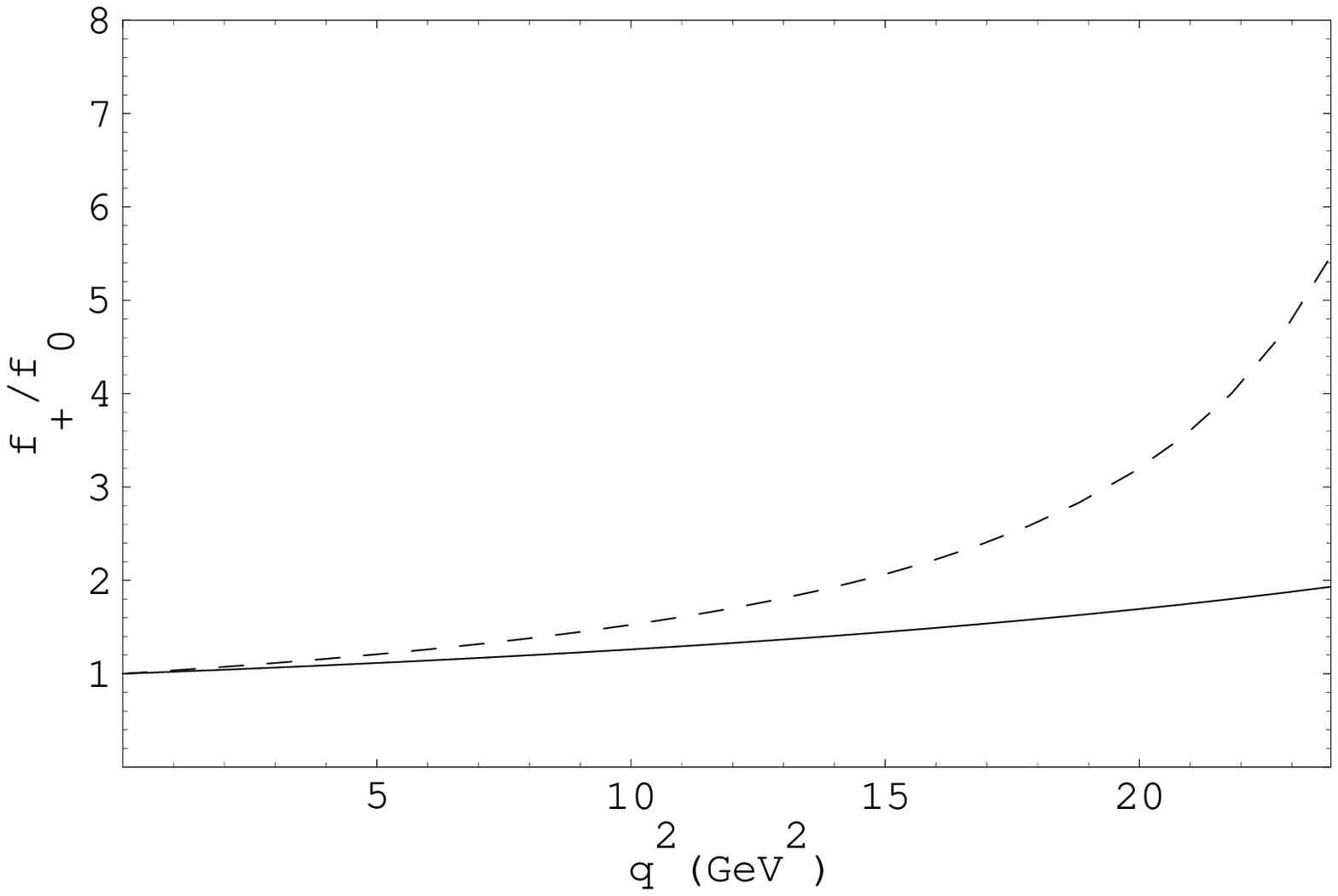}{(b)}

\PICLI{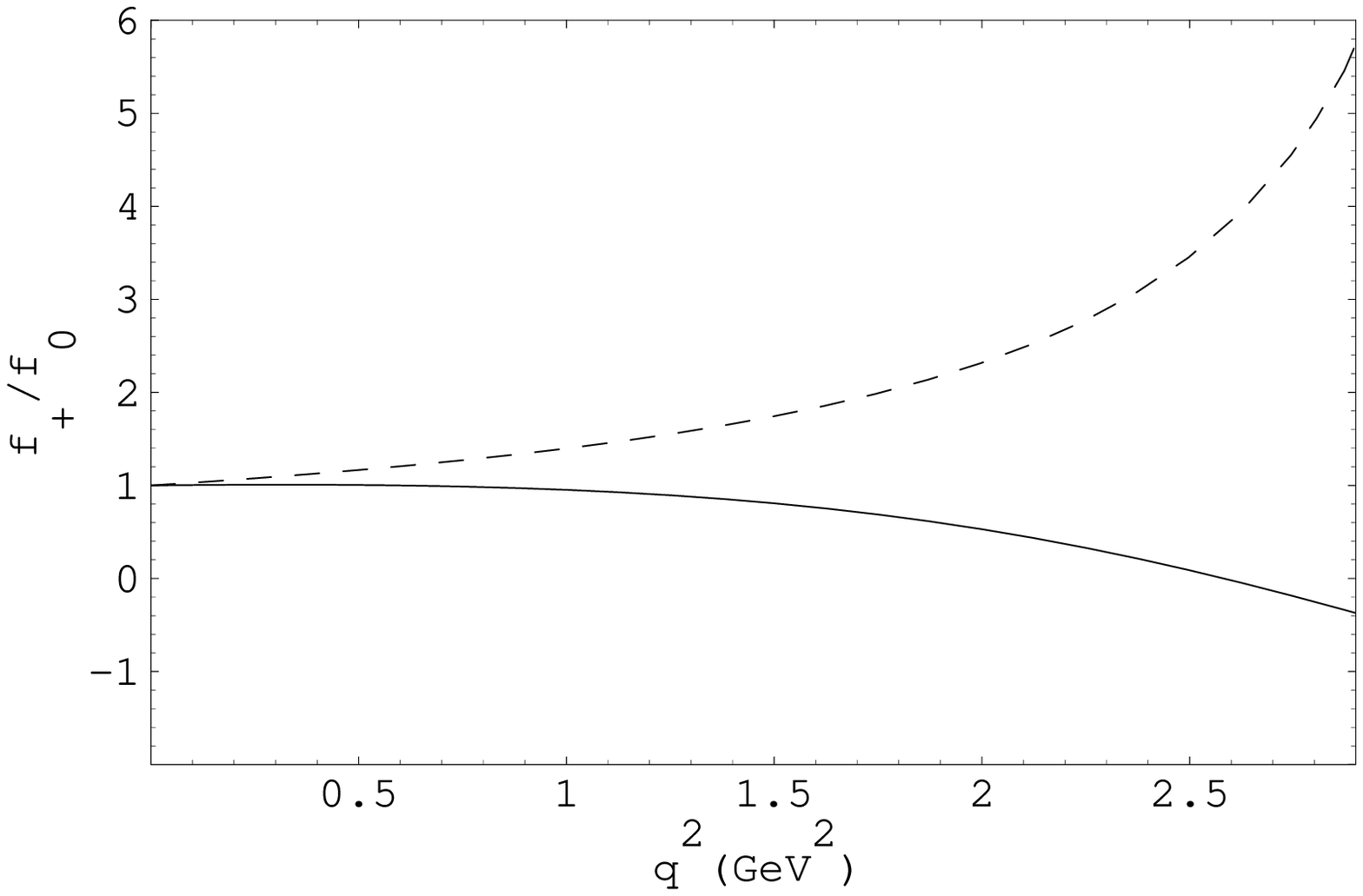}{(c)}

\PICRI{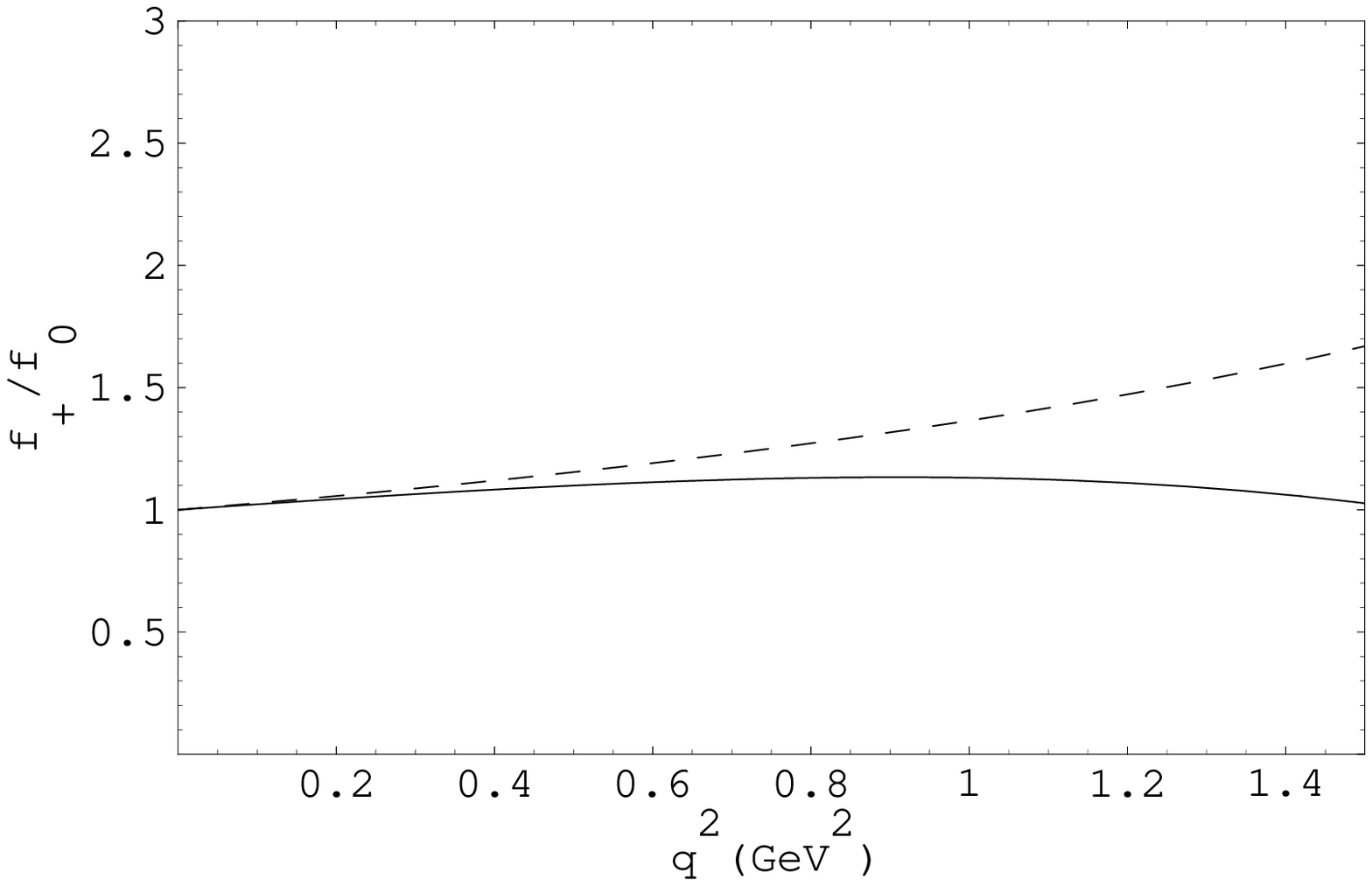}{(d)}

\vspace{-2cm}
\centerline{
\parbox{12cm}{
\small
\baselineskip=1.0pt
Fig.3. Results for the ratio $f_+(q^2)/f_0(q^2)$. (a), (b), (c), (d) are for 
$B\to \pi l\nu$, $B_s \to Kl\nu$, $D\to \pi l \nu$, $D\to Kl\nu$ respectively. 
Solid lines are our sum rule results, while the dashed lines are produced 
from the LEET relation (\ref{leet1}).
} }

\newpage

\mbox{}
{\vspace{1.2cm}}

\PICLI{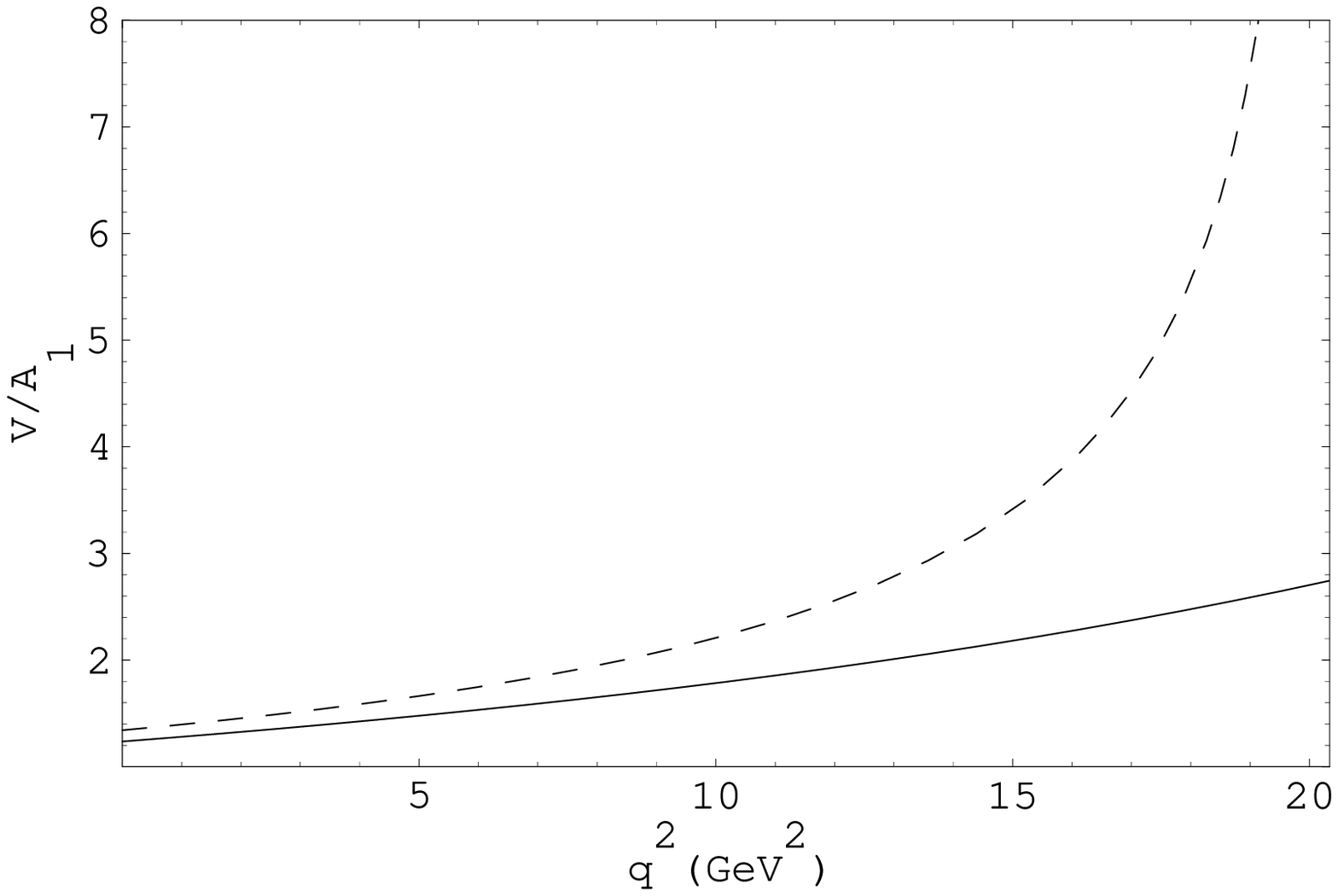}{(a)}

\PICRI{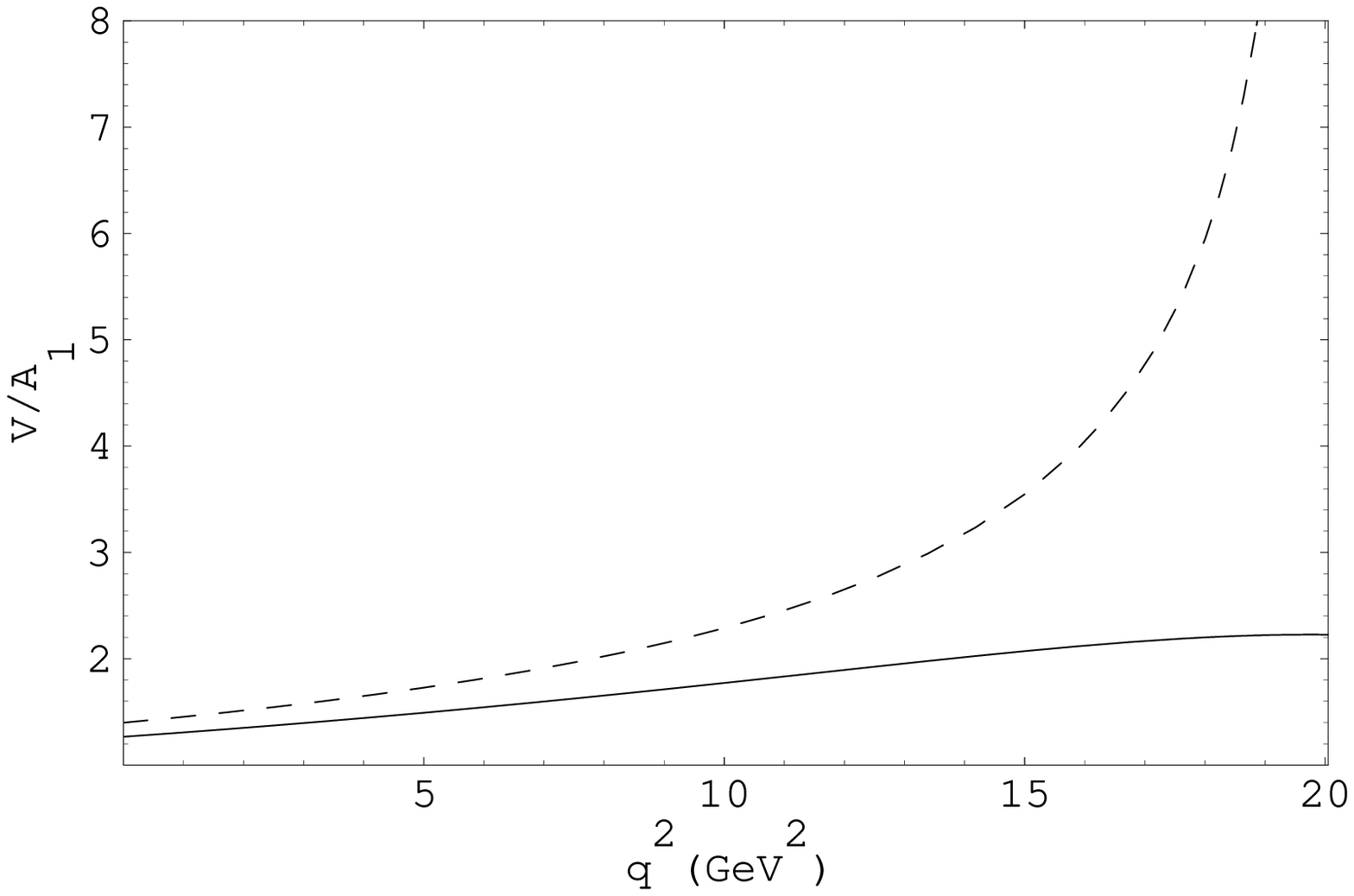}{(b)}

\PICLI{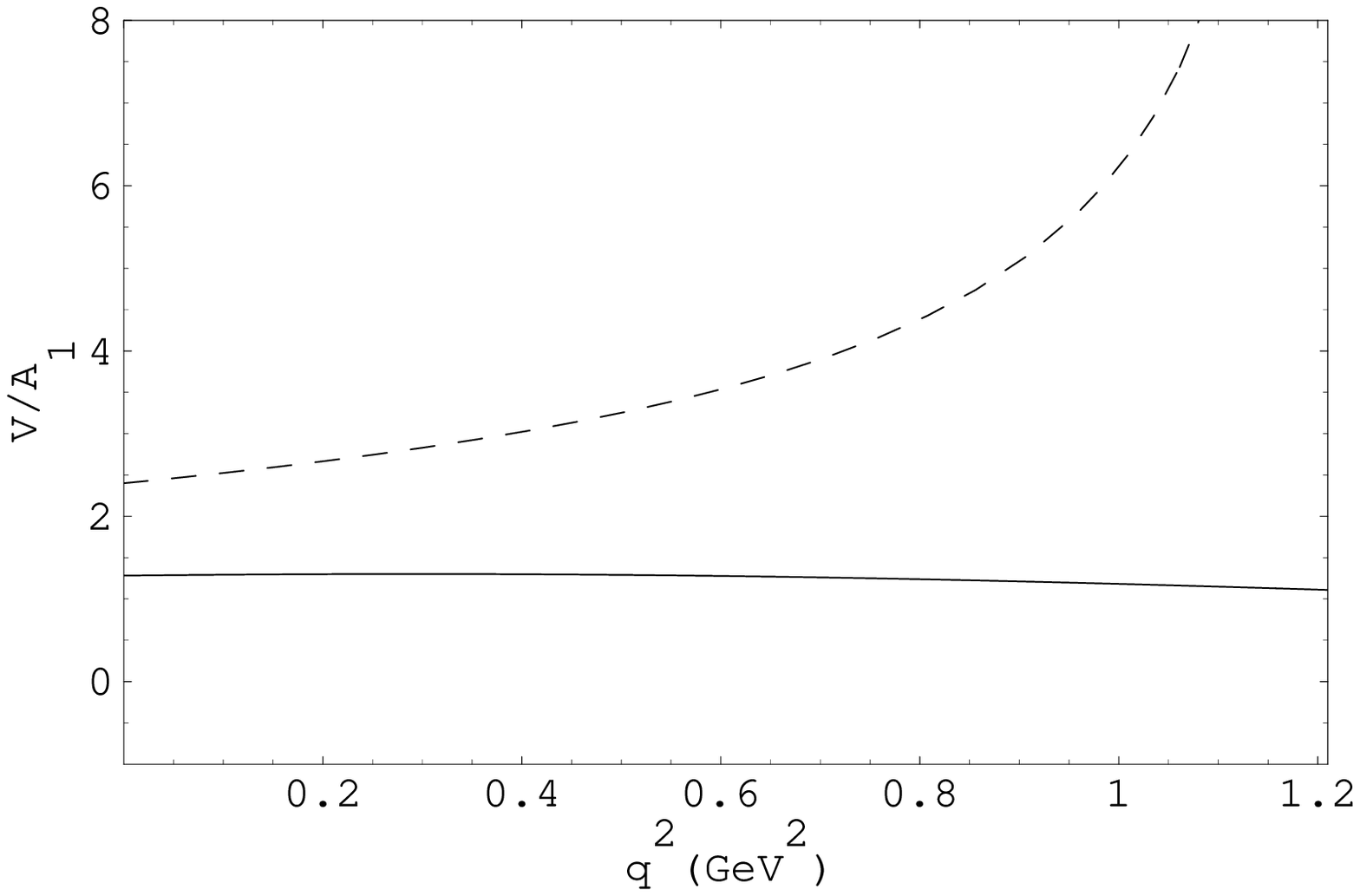}{(c)}

\PICRI{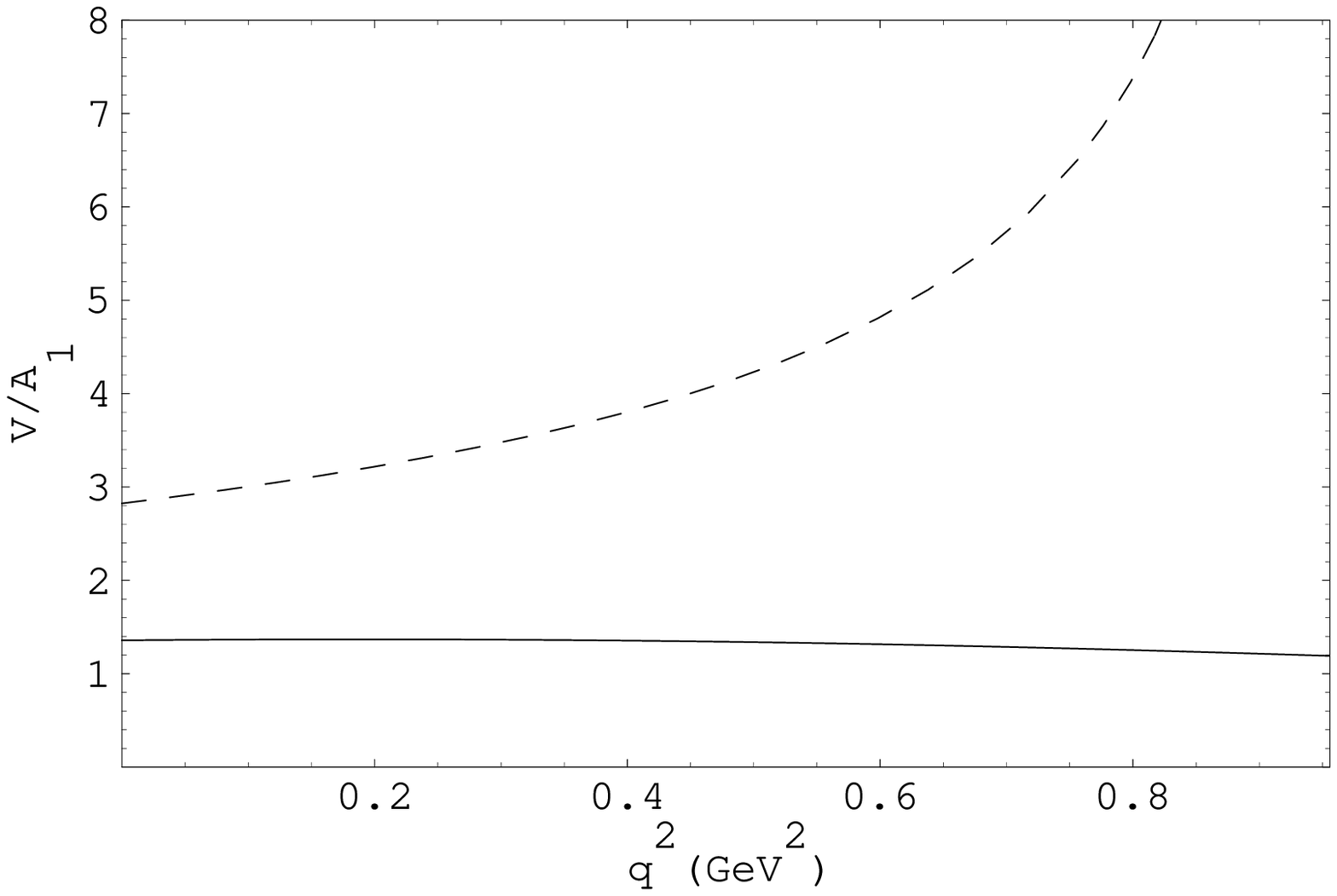}{(d)}

\vspace{-2cm}
\centerline{
\parbox{12cm}{
\small
\baselineskip=1.0pt
Fig.4. Results for the ratio $V(q^2)/A_1(q^2)$. (a), (b), (c), (d) are for 
$B\to \rho l\nu$, $B_s \to K^* l \nu$, $D\to \rho l \nu$, $D\to K^* l\nu$ 
respectively. 
Solid lines are our sum rule results, while the dashed lines are produced 
from the LEET relation (\ref{leet2}).
} }

\newpage

\mbox{}
{\vspace{1.2cm}}

\PICLI{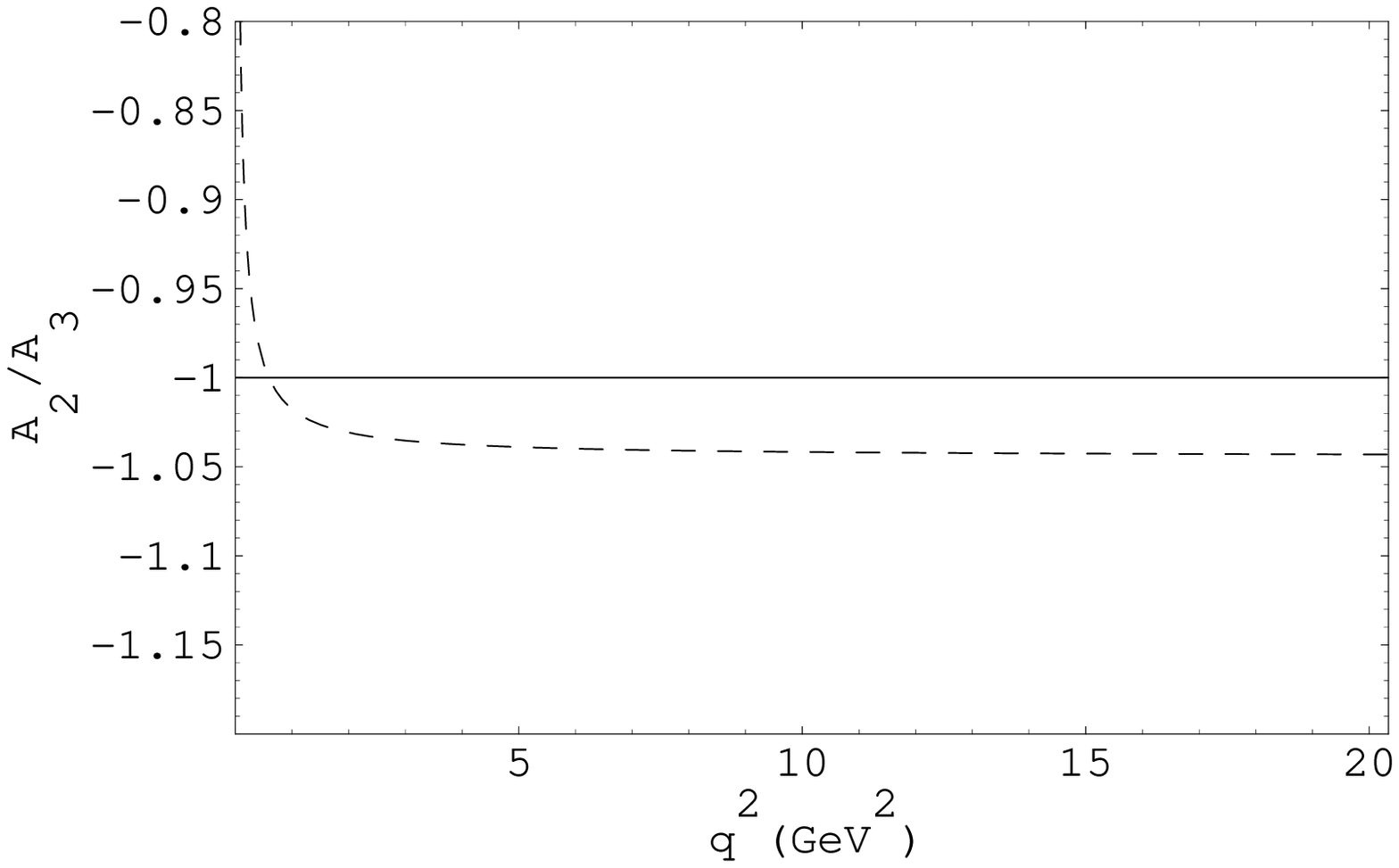}{(a)}

\PICRI{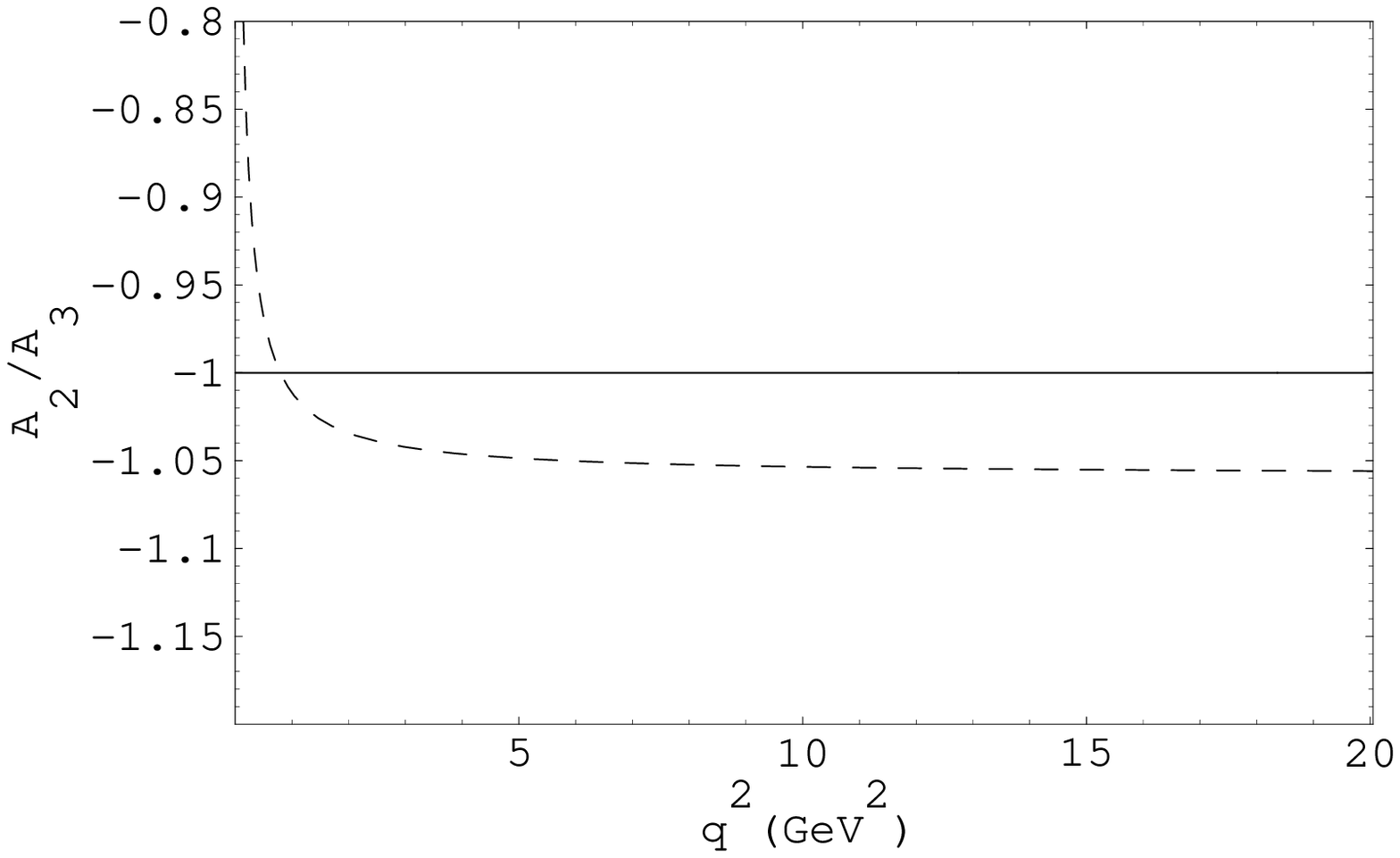}{(b)}

\PICLI{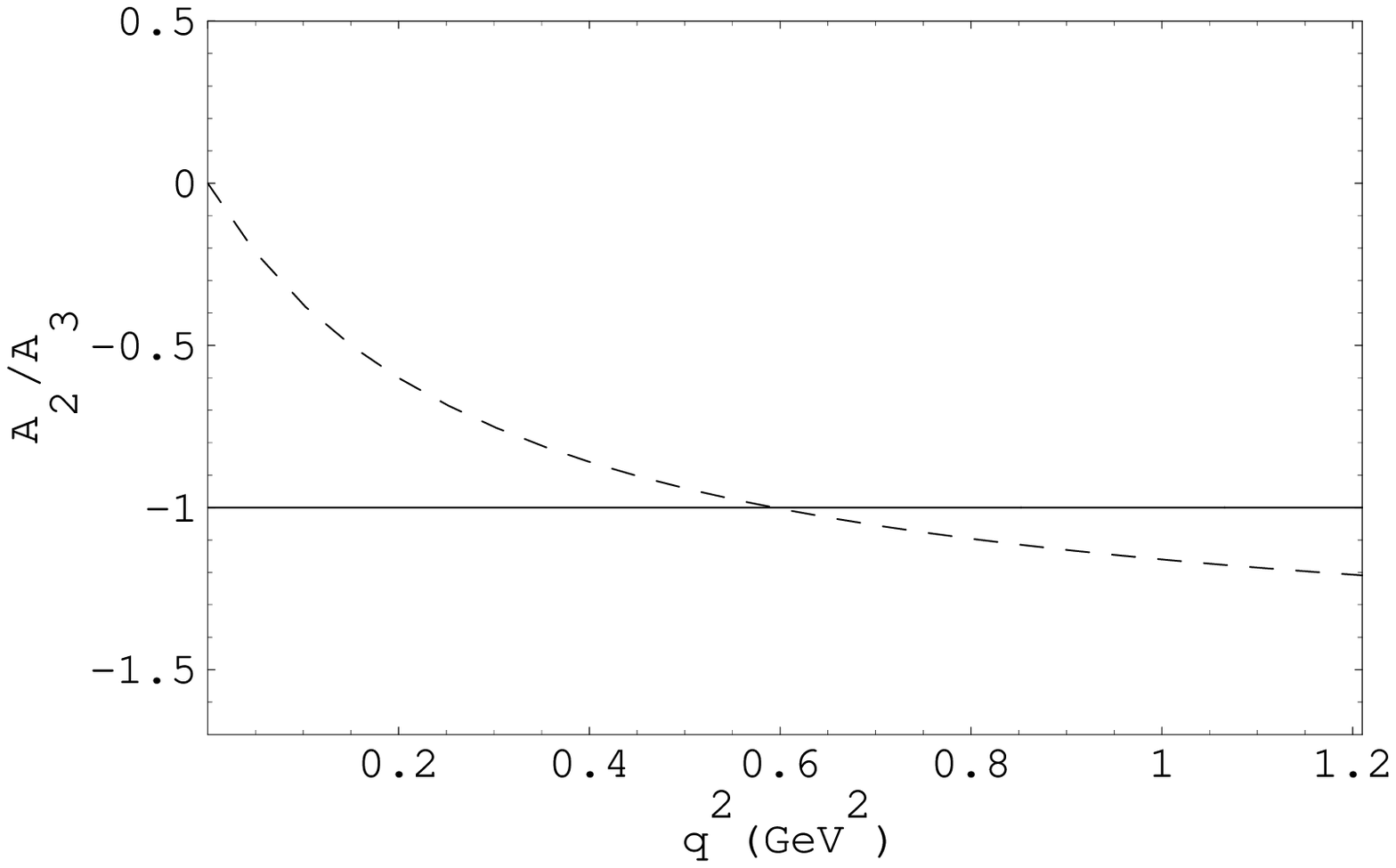}{(c)}

\PICRI{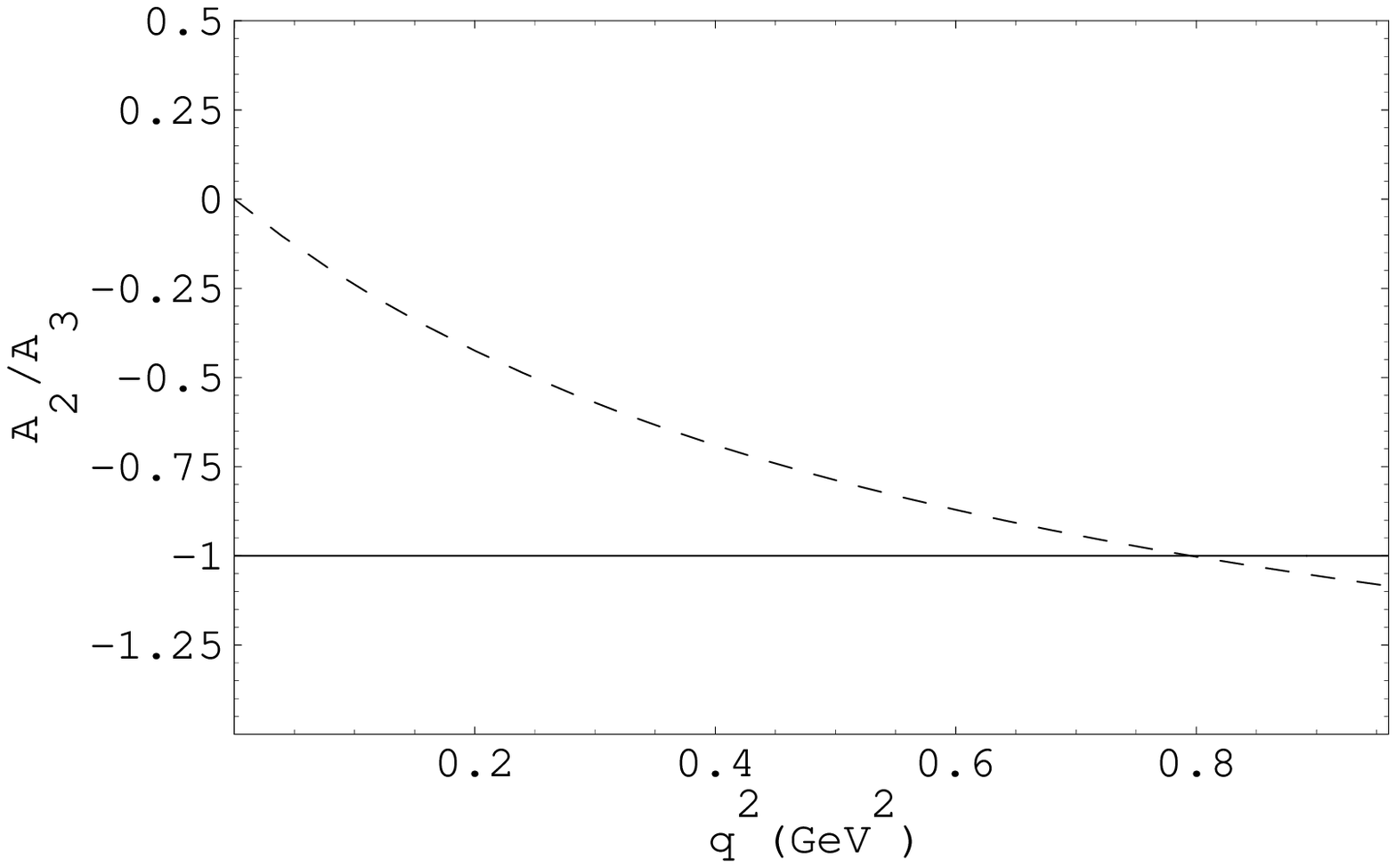}{(d)}

\vspace{-2cm}
\centerline{
\parbox{12cm}{
\small
\baselineskip=1.0pt
Fig.5. Results for the ratio $A_2(q^2)/A_3(q^2)$. (a), (b), (c), (d) are for 
$B\to \rho l\nu$, $B_s \to K^* l \nu$, $D\to \rho l \nu$, $D\to K^* l\nu$ 
respectively. 
Solid lines are our sum rule results, while the dashed lines are produced 
from the LEET relation (\ref{leet3}).
} }


\begin{references}

\bibitem{ape} C. R. Allton {\it et al}., Phys. Lett. B {\bf 345}, 513 (1995).
\bibitem{wuppertal} S. Gusken {\it et al}., Nucl. Phys. (Proc. Suppl.) {\bf 47}, 
    485 (1996).
\bibitem{ukqcd} J. Nieves {\it et al}., Nucl. Phys. (Proc. Suppl.) {\bf 42}, 
    431 (1995).
\bibitem{elc} A. Abada {\it et al}., Nucl. Phys. B {\bf 416}, 675 (1994).
\bibitem{lat9910021} A. Abada, D. Becirevic, Ph. Boucaud, J. P. Leroy, 
    V. Lubicz, G. Martinelli, F. Mescia, Nucl. Phys. Proc. Suppl. {\bf 83}, 268 (2000).
\bibitem{lat9710057} J. M. Flynn, C. T. Sachrajda, Adv. Ser. Direct. High Energy 
     Phys. {\bf 15}, 402 (1998). 
\bibitem{lat98} L. Del Debbio, J. M. Flynn, L. Lellouch, J. Nieves 
    (UKQCD Collaboration), Phys. Lett. B {\bf 416}, 392 (1998).
\bibitem{ph0001113} D. Melikhov, B. Stech, Phys. Rev. D {\bf 62}, 014006 (2000).
\bibitem{Jaus96} W. Jaus, Phys. Rev. D {\bf 41}, 3394 (1990); {\it ibid} D {\bf 53}, 
    1349 (1996). 
\bibitem{wsb85} M. Wirbel, B. Stech, M. Bauer, Z. Phys. C {\bf 29}, 637 (1985).  
\bibitem{ph9807223} M. Beyer, D. Melikhov, Phys. Lett. B {\bf 436}, 344 (1998).
\bibitem{isgw2} N. Isgur, D. Scora, B. Grinstein, M. Wise, Phys. Rev. D {\bf 39}, 
     799 (1989); D. Scora, N. Isgur, Phys. Rev. D {\bf 52}, 2783 (1995).
\bibitem{pvesrb} P. Ball and V. M. Braun, Phys. Rev. D {\bf 58}, 094016 (1998).
\bibitem{var} V. M. Belyaev, A. Khodjamirian and R. R\"uckl, Z. Phys. C {\bf 60}, 349 (1993).
\bibitem{ar} A. Khodjamirian and R. R\"uckl, WUE-ITP-97-049, MPI-PhT/97-85.
\bibitem{arsc} A. Khodjamirian, R. R\"uckl, S. Weinzierl, C. W. Winhart and O. Yakovlev,
    Phys. Rev. D {\bf 62}, 114002 (2000).
\bibitem{ph9805422} P. Ball and V. M. Braun, Phys. Rev. D {\bf 58}, 094016 (1998).
\bibitem{ph9802394} P. Ball, JHEP 9809, 005 (1998). 
\bibitem{ph0001297} A. Khodjamirian, R. R$\ddot{u}$ckl, S. Weinzierl, C. W. Winhart, 
    O. Yakovlev, Phys. Rev. D {\bf 62}, 114002 (2000).
\bibitem{ph9701238} P. Ball, V. M. Braun, Phys. Rev. D {\bf 55}, 5561 (1997).
\bibitem{sr9193} P. Ball, V. Braun, H. Dosch, Phys. Lett. B {\bf 273}, 316 (1991);
     P. Ball, V. Braun, H. Dosch, Phys. Rev. D {\bf 44}, 3567 (1991); 
     P. Ball, Phys. Rev. D {\bf 48}, 3190 (1993).
\bibitem{gzmy} G. Burdman, Z. Ligeti, M. Neubert and Y. Nir, Phys. Rev. D {\bf 49}, 2331 (1994).
\bibitem{hly} C. S. Huang, C. Liu and C. T. Yan, Phys.Rev. D {\bf 62}, 054019 (2000).
\bibitem{bpi} W. Y. Wang, Y. L. Wu, Phys. Lett. B {\bf 515}, 57 (2001).
\bibitem{brho} W. Y. Wang, Y. L. Wu, Phys. Lett. B {\bf 519}, 219 (2001); 
      hep-ph/0106208 (corrected version).
\bibitem{ylw} Y. L. Wu, Mod. Phys. Lett. A {\bf 8}, 819 (1993).
\bibitem{wwy} W. Y. Wang, Y. L. Wu and Y. A. Yan, Int. J. Mod. Phys. A {\bf 15}, 1817 (2000).
\bibitem{ww} W. Y. Wang and Y. L. Wu, Int. J. Mod. Phys. A {\bf 16}, 377 (2001).
\bibitem{va} V. L. Chernyak and A. R. Zhitnitsky, Phys. Rep. {\bf 112}, 173 (1984).
\bibitem{bf} V. M. Braun and I. B. Filyanov, Z. Phys. C {\bf 48}, 239 (1990).
\bibitem{vvar} V. M. Belyaev, V. M. Braun, A. Khodjamirian and R. Rckl, Phys.Rev. 
     D {\bf 51}, 6177 (1995).
\bibitem{pb} P. Ball, JHEP {\bf 01}, 010 (1999).
\bibitem{vi} V. M. Braun and I. B. Filyanov, Z. Phys, C {\bf 44}, 157 (1989).
\bibitem{cleo} J. P. Alexander et. al.(CLEO Collab.), Phys. Rev. Lett. {\bf 77}, 5000 (1996).
\bibitem{pdg} Particle Data Group, Phys. Rev. D {\bf 54}, 1 (1996).
\bibitem{pvda} P. Ball and V. M. Braun, Phys. Rev. D {\bf 54}, 2182 (1996).
\bibitem{pvmisu} P. Ball and V. M. Braun, Phys. Rev. D {\bf 55}, 5561 (1997).
\bibitem{mab} M. A. Benitez et. al., Phys. Lett. B {\bf 239}, 1 (1990).
\bibitem{lm} L. Montanet et. al., Phys. Rev. D {\bf 50}, 1173 (1994).
\bibitem{zfxt} Z. H. Li, F. Y. Liang, X. Y. Wu and T. Huang, Phys. Rev. D {\bf 64}, 057901 (2001).  
\bibitem{beatrice} M. Adamovich {\it et al}., Eur. Phys. J. C {\bf 6}, 35 (1999). 
\bibitem{e791} E. M. Aitala {\it et al}., Phys. Rev. Lett. {\bf 80}, 1393 (1998).
\bibitem{e687} P. L. Frabetti {\it et al}., Phys. Lett. B {\bf 307}, 262 (1993).
\bibitem{e653} K. Kodama {\it et al}., Phys. Lett. B {\bf 274}, 246 (1992).
\bibitem{e691} J. C. Anjos {\it et al}., Phys. Rev. Lett. {\bf 65}, 2630 (1990).
\bibitem{dugan} M. J. Dugan, B. Grinstein, Phys. Lett. B {\bf 255}, 583 (1991).
\bibitem{uag} U. Aglietti, Phys. Lett. B {\bf 292}, 424 (1992); 
       U. Aglietti, G. Gorbo, Phys. Lett. B {\bf 431}, 166 (1998).
\bibitem{jalo} J. Charles, A. Le Yaouanc, L. Oliver, O. Pene, J.-C. Raynal, 
     Phys. Rev. D {\bf 60}, 014001 (1999).  
\bibitem{mth} M. Beneke, Th. Feldmann, Nucl. Phys. B {\bf 592}, 3 (2001).  
\bibitem{ph0107065} D. Ebert, R. N. Faustov, V. O. Galkin, 
      Phys. Rev. D {\bf 64}, 094022 (2001). 
\bibitem{epjc} Particle Data Group, Eur. Phys. J. C {\bf 15}, 1 (2000). 
\bibitem{wa82} M. Adamovich {\it et al}., Phys. Lett. B {\bf 268}, 142 (1991).

\end{references}
\end{document}